\documentclass[cernpreprint,atlasdraft=false, USenglish, texmf, orcidlogo]{atlasdoc}
\usepackage[subfigure]{atlaspackage}

\usepackage{atlasbiblatex}

\usepackage{atlasphysics}
\usepackage{atlasjetetmiss}

\graphicspath{{logos/}{figures/}}

\usepackage{ANA-EXOT-2019-01-PAPER-defs}
\usepackage{comment}

\AtlasTitle{Search for low-mass resonances decaying into two jets and produced in association with a photon or a jet at $\sqrt{s}=13$ TeV with the ATLAS detector}

\AtlasAbstract{%
A search is performed for localized excesses in the low-mass dijet invariant mass distribution, targeting a hypothetical new particle decaying into two jets and produced in association with either a high transverse momentum photon or a jet.
The search uses the full Run~2 data sample from LHC proton--proton collisions collected by the ATLAS experiment at a center-of-mass energy of 13~TeV during 2015--2018.
Two variants of the search are presented for each type of initial-state radiation: one that makes no jet flavor requirements and one that requires both of the jets to have been identified as containing $b$-hadrons.
No excess is observed relative  to the Standard Model prediction, and the data are used to set upper limits on the production cross-section for a benchmark $Z'$ model and,
separately, for generic, beyond the Standard Model scenarios which might produce a Gaussian-shaped contribution to dijet invariant mass distributions. The results extend the current constraints on dijet resonances to the mass range between 200 and 650~GeV.
}

\AtlasRefCode{EXOT-2019-01}

\PreprintIdNumber{CERN-EP-2024-044}

\AtlasJournalRef{Phys.\ Rev.\ D 110 (2024) 3, 032002}
\AtlasDOI{10.1103/PhysRevD.110.032002}

\hypersetup{pdftitle={ATLAS document},pdfauthor={The ATLAS Collaboration}}

\begin{document}

\maketitle

\section{Introduction} %

% The next lines are included from the .//sections/intro.tex input file
The Standard Model of particle physics (SM) successfully describes a wide range of phenomena across many orders of magnitude of energy at colliders, astrophysical observatories, and other experiments~\cite{Zyla:2020zbs}.
Nevertheless it fails to account for many observed phenomena,
such as the presence of dark matter (DM) \cite{Bertone:2004pz,Bergstrom:2000pn,Clowe:2006eq} and the imbalance of matter over antimatter in the observable universe~\cite{Canetti:2012zc}.
While there are a wide range of proposed models of physics beyond the SM (BSM) to explain these effects~\cite{Martin:1997ns,Aguilar-Saavedra:2009xmz,Giudice:2017fmj,Strassler:2006im,Fan:2011yu,Ghosh:2012ep,Konar:2010bi,Florez2019,Abdallah:2015ter,Bernal:2017kxu,Arias:2012az,Kahlhoefer:2015bea},
this paper focuses on a search for a hadronically decaying resonance.

\Zp~bosons are hypothetical spin-1 vector bosons that are singlets under electric and color charge~\cite{Zyla:2020zbs}, predicted in a variety of BSM models as a potential mediator to dark matter particles~\cite{Abercrombie:2015wmb,Abdallah:2015ter,Kahlhoefer:2017dnp,Fairbairn:2016iuf,Aaboud:2019yqu,Buchmueller:2013dya,Kahlhoefer:2015bea}.
Direct \Zp~searches have shown to be an effective way of constraining BSM models~\cite{Kahlhoefer:2017dnp,Fairbairn:2016iuf,Aaboud:2019yqu}.
Searches at the Large Hadron Collider (LHC) for \Zp~decays into pairs of charged leptons set strong constraints~\cite{ATL-PHYS-PUB-2023-018,,Aaboud:2019yqu},
but may be eluded by choosing the couplings of the \Zp~such that it is \emph{leptophobic}, meaning it does not couple to SM leptons.
However, constraints from searches for hadronically decaying \Zp~resonances cannot be avoided in this way, since the same %
$\Zp\text{-}q\text{-}q$
decay vertex
would also be responsible for \Zp~production at the LHC for any kind of resonance or %
search for topologies with a single reconstructed object recoiling against a large missing transverse energy.
Additionally, the %
$\Zp\text{-}q\text{-}q$
vertex would be necessary for the dark matter self-annihilation cross-section that drives the relic-density for weakly interacting massive particles~\cite{Bertone:2004pz}.

The ATLAS and CMS Collaborations at the LHC have published several searches that are sensitive to hadronic \Zp~production using a variety of strategies.
Searches for excesses in the invariant mass of the two jets with the highest transverse momentum (\pt) in the event constrain \Zp~production at high masses
~\cite{EXOT-2011-07,EXOT-2013-11,CMS-EXO-10-010,CMS-EXO-11-015,CMS-EXO-11-094,CMS-EXO-12-016,CMS-EXO-12-059,CMS-EXO-14-005,EXOT-2015-02,CMS-EXO-15-001,CMS-EXO-16-032,CMS-EXO-16-056,CMS-EXO-16-057,CMS-EXO-19-012,CMS-EXO-20-008}.
The lower reach of these searches are limited by the bandwidth available to single- and multijet triggers, and so different strategies must be used to gain sensitivity to \Zp~masses below around twice the \pt threshold where these triggers become fully efficient, around 1~\TeV.
One strategy is to record only minimal information, but at a higher rate than the standard triggers, enabling lower trigger thresholds~\cite{EXOT-2016-20,CMS-EXO-14-005}.
A complementary approach, explored in this paper, relies on the production of high-\pt~initial-state-radiation (ISR), such as a jet or a photon, which recoils against the \Zp, enabling access to lower dijet masses without trigger bias.
This type of search was performed by the ATLAS and CMS Collaborations in topologies where the two \Zp~decay products can be resolved as individual jets~\cite{EXOT-2018-05,CMS-EXO-19-004}
and the case where the decay products are sufficiently boosted to be reconstructed into a single large-radius jet~\cite{CMS-EXO-17-027,CMS-EXO-18-012,CMS-EXO-17-001,CMS-EXO-16-030,EXOT-2017-01}.

This paper explores the resolved case, where the resonance decay products are reconstructed into two jets, covering a resonance mass range of 200~\GeV~to 650~\GeV.
In total, four different channels are studied, based on the type of ISR particle and the flavor composition of the \Zp~decay products.
The first channel, \ajj, selects events in which the ISR is a photon and applies an inclusive selection on the jets that form the $Z'$.
The second channel, \abb, focuses on the case where the ISR is a photon and both jets are identified as containing $b$-hadrons ($b$-tagged).
The third and fourth channels focus on the case where the ISR is a jet, both in the inclusive case (\jjj), and the case where both jets from the candidate resonance decay are $b$-tagged (\jbb).
While $b$-tagging requirements impose assumptions on the \Zp~decay products, they also provide significant background suppression, resulting in a more powerful search for some models.
The cross-sections for the \jjj~and \jbb~processes are higher than for \ajj~and \abb.
However, the \jjj~ channel has the additional challenge of identifying which jets might correspond to the \Zp~decay products, and which correspond to the ISR\@.
Furthermore, different trigger thresholds apply for recorded events triggered by a photon or a jet. Hence a different kinematic acceptance is imposed on the ISR object for the various channels.
The two channels where the ISR is a photon are collectively called the \textit{photon} channels, while the two where the ISR is a jet are referred to as the \textit{trijet} channels.

In addition to the increased luminosity, this analysis features several improvements over the previous analysis targetting the same signature~\cite{EXOT-2018-05},
including a better $b$-tagging algorithm that reduces the background and the inclusion of the trijet channels.
The dominant backgrounds for this search are multijet and single-photon production, which are estimated from data using a functional form fit to the \mjj~distribution.
For each channel, estimates for background and signal are combined into a likelihood function, where the signal yields are controlled by a free multiplicative parameter.
The signal is interpreted as either a \Zp~boson or a generic narrow resonance producing a Gaussian-shaped bump in the measured mass spectrum.

The paper is organized as follows. The ATLAS detector at the LHC and the data and simulated samples used in this search are described in Sections~\ref{sec:atlas}--~\ref{sec:samples}.
The object and event selection for the four channels are summarized in Section~\ref{sec:eventSelection}.
The methodology of the search is introduced in Section~\ref{sec:methodology} and the systematic uncertainties are discussed in Section~\ref{sec:systematics}.
The results are presented in Section~\ref{sec:results} and conclusions are drawn in Section~\ref{sec:conclusion}.

% End of text imported from the .//sections/intro.tex input file

%

%
\section{The ATLAS detector}
\label{sec:atlas}

% The next lines are included from the .//sections/atlas.tex input file
\newcommand{\AtlasCoordFootnote}{%
ATLAS uses a right-handed coordinate system with its origin at the nominal interaction point (IP) in the center of the detector and the \(z\)-axis along the beam pipe. The \(x\)-axis points from the IP to the center of the LHC ring, and the \(y\)-axis points upwards. Cylindrical coordinates \((r,\phi)\) are used in the transverse plane, \(\phi\) being the azimuthal angle around the \(z\)-axis. The pseudorapidity is defined in terms of the polar angle \(\theta\) as \(\eta = -\ln \tan(\theta/2)\). Angular distance is measured in units of \(\Delta R \equiv \sqrt{(\Delta y)^{2} + (\Delta\phi)^{2}}\), where $y=(1/2)[(E+p_z)/(E-p_z)]$ is the object's rapidity defined by its energy and longitudinal momentum.}

The ATLAS detector~\cite{PERF-2007-01} at the LHC covers nearly the entire solid angle around the collision point.\footnote{\AtlasCoordFootnote}
It consists of an inner tracking detector surrounded by a thin superconducting solenoid, electromagnetic and hadron calorimeters,
and a muon spectrometer incorporating three large superconducting air-core toroidal magnets.

The inner-detector system is immersed in a \SI{2}{\tesla} axial magnetic field
and provides charged-particle tracking in the range of \(|\eta| < 2.5\).
The high-granularity silicon pixel detector covers the vertex region and typically provides four measurements per track,
the first hit normally being in the insertable B-layer %
installed before \RunTwo~\cite{ATLAS-TDR-19,PIX-2018-001}.
It is followed by the silicon microstrip tracker%
, which usually provides eight measurements per track.
These silicon detectors are complemented by the transition radiation tracker (TRT),
which enables radially extended track reconstruction up to \(|\eta| = 2.0\).
The TRT also provides electron identification information
based on the fraction of hits (typically 30 in total) above a higher energy-deposit threshold corresponding to transition radiation.

The calorimeter system covers the pseudorapidity range \(|\eta| < 4.9\).
Within the region \(|\eta|< 3.2\), electromagnetic calorimetry is provided by barrel and
endcap high-granularity lead/liquid-argon (LAr) calorimeters,
with an additional thin LAr presampler covering \(|\eta| < 1.8\)
to correct for energy loss in material upstream of the calorimeters.
Hadron calorimetry is provided by the steel/scintillator-tile calorimeter,
segmented into three barrel structures within \(|\eta| < 1.7\), and two copper/LAr hadron endcap calorimeters.
The solid angle coverage is completed with forward copper/LAr and tungsten/LAr calorimeter modules
optimized for electromagnetic and hadronic energy measurements respectively.

The muon spectrometer %
comprises separate trigger and
high-precision tracking chambers measuring the deflection of muons in a magnetic field generated by the superconducting air-core toroidal magnets.
The field integral of the toroids ranges between \num{2.0} and \SI{6.0}{\tesla\metre}
across most of the detector.
Three layers of precision chambers, each consisting of layers of monitored drift tubes, cover the region \(|\eta| < 2.7\),
complemented by cathode-strip chambers in the forward region, where the background is highest.
The muon trigger system covers the range \(|\eta| < 2.4\) with resistive-plate chambers in the barrel, and thin-gap chambers in the endcap regions.

Interesting events are selected by the first-level trigger system implemented in custom hardware,
followed by selections made by algorithms implemented in software in the high-level trigger~\cite{TRIG-2016-01}.
The first-level trigger accepts events from the \SI{40}{\MHz} bunch crossings at a rate below \SI{100}{\kHz},
which the high-level trigger further reduces to record events to disk at about \SI{1}{\kHz}.

An extensive software suite~\cite{ATL-SOFT-PUB-2021-001} is used in data simulation, in the reconstruction and analysis of real
and simulated data, in detector operations, and in the trigger and data acquisition systems of the experiment.

% End of text imported from the .//sections/atlas.tex input file

\section{Data and Monte Carlo simulated samples}
\label{sec:samples}

% The next lines are included from the .//sections/samples.tex input file

This analysis is performed using data from proton--proton ($pp$) collisions at $\sqrt{s}=13~\TeV$ at the LHC, collected during 2015--2018 with the ATLAS detector.
The total integrated luminosity of this data sample is $140~\ifb$ \cite{DAPR-2021-01},
obtained using the LUCID-2 detector~\cite{LUCID2} for the primary luminosity measurements.
Due to the high instantaneous luminosity and the large total inelastic $pp$ cross-section, there are, on average, 33.7 %
collisions in each bunch crossing.
The additional collisions not selected as the hard-scatter process are referred to as \pileup collisions.
Data in this analysis are required to satisfy standard quality requirements~\cite{DAPR-2018-01}.

Samples of Monte Carlo (MC) simulated signal and background %
events are used in this analysis for optimization, estimate of possible signal contributions, and validation of background estimation strategies.

The main benchmark model is a simplified leptophobic \Zp~axial-vector mediator signal, which is used to test the sensitivity to potential resonant signals.
The choice of this benchmark signal is motivated by the role that dijet resonance searches play in constraining possible interactions between dark matter and SM particles\@.
This simplified model considers an $s$-channel mediator with axial-vector interactions, as described in Refs.~\cite{Albert:2017onk,Abercrombie:2015wmb}.
Two relevant parameters of this model in the dijet search are  the \Zp~mass, and the coupling of the \Zp~to quarks ($g_q$), which affect the intrinsic width of the resonance.
The mass of the dark matter fermion is set to be much heavier than the \Zp~mass, such that the decay width into dark matter is zero.
In all channels, signal samples are generated for \Zp~masses between 200~\GeV and 650~\GeV, with the coupling of the mediator to non-top-quarks set to 0.2, regardless of the quark flavor.
Such a choice for the coupling value is a benchmark model already used in other analyses~\cite{EXOT-2018-05}.
Also, given the \Zp~intrinsic width for $g_q=0.2$ is much smaller than the experimental resolution, such a sample is adequate to also model the shape of signals with slightly different or smaller $g_q$ values.
Signal samples are generated separately for a \Zp~decaying into $u$,$d$,$s$ and $c$ quark flavors and \Zp~decaying into $b$-quarks, in order to enhance the signal statistical precision for the $b$-tagged channels.

The signal samples are simulated using the \MADGRAPH generator~\cite{Alwall:2014hca} at leading-order (LO) in perturbative quantum chromodynamics (QCD) with the \NNPDF[2.3lo] parton distribution function (PDF) set~\cite{Ball:2012cx}. The events are interfaced to \PYTHIA{8.244}~\cite{Sjostrand:2014zea,Sjostrand:2006za} to model the parton shower, hadronization, and underlying event, with parameter values set according to the A14 tune~\cite{ATL-PHYS-PUB-2014-021}. EvtGen~\cite{Lange:2001uf} is used to model decays of heavy flavor hadrons.
To further increase statistical precision in the \ajj~channel, a generator-level filter is applied on the photon \pt.
Two sets of samples are produced: one set with this filter defined to be $p_\mathrm{T,\gamma} \geq 130$~\GeV and a second with the filter at $p_\mathrm{T,\gamma} \geq 65$~\GeV.
For the trijet channels, a generator-level jet filter of 370~\GeV is applied on the leading-\pt $R=0.4$ jet, which is fully efficient for the reconstructed jet \pt threshold.
The generator-level filters are chosen based on the triggers used in the two different analyses, to ensure high statistical precision for each mass point without introducing kinematic biases.

While the final background estimate is performed using a data-driven approach, background multijet and photon-plus-jets samples are used to optimize the event selection and validate the background estimation strategy.
For the photon channels, prompt single-photon production is simulated with the \SHERPA[2.2]~\cite{Bothmann:2019yzt} generator.
In this arrangement and thanks to the Comix~\cite{Gleisberg:2008fv} and \OPENLOOPS~\cite{Buccioni:2019sur,Cascioli:2011va,Denner:2016kdg} libraries, matrix elements for up to two partons are calculated to next-to-leading-order (NLO) accuracy, while matrix elements for up to four partons are calculated at LO in QCD.
They are matched with the \SHERPA parton shower~\cite{Schumann:2007mg} using the \MEPSatNLO prescription~\cite{Hoeche:2011fd,Hoeche:2012yf,Catani:2001cc,Hoeche:2009rj} with a dynamic merging scale cut~\cite{Siegert:2016bre} of \qty{20}{\GeV}.
Photons are required to be isolated according to a smooth-cone isolation criterion~\cite{Frixione:1998jh}.
Samples are simulated using the \NNPDF[3.0nnlo] PDF set~\cite{Ball:2014uwa}, along with the dedicated set of tuned parton-shower parameters developed by the \SHERPA authors.

For the trijet channels, background samples of simulated multijet processes are generated using \PYTHIA~8.230 with the $2\rightarrow 2$ matrix element,
and additional jets are produced through the parton shower.
Events are simulated using the A14 tune~\cite{ATL-PHYS-PUB-2014-021}, the Lund string hadronization model and the \NNPDF[2.3lo] PDF set.
The \PYTHIA parton shower algorithm uses a dipole-style \pt-ordered evolution, and its renormalisation and factorisation scales are set to the geometric mean of the squared transverse masses of the outgoing particles.

The generated background and signal events are passed through a detailed detector simulation~\cite{SOFT-2010-01} based on \GEANT~\cite{Agostinelli:2002hh}.
Additional minimum-bias interactions simulated using \PYTHIA{8}~with the A3 tune~\cite{ATL-PHYS-PUB-2016-017} and the \NNPDF[2.3lo] PDF set~\cite{Ball:2012cx} are overlayed to the backgrond and signal events, to model \pileup interactions.
The distribution of the average number of \pileup interactions in simulation is reweighted during data analysis to match that observed in \RunTwo data.

% End of text imported from the .//sections/samples.tex input file

\FloatBarrier

\section{Event reconstruction and selection}
\label{sec:eventSelection}

% The next lines are included from the .//sections/eventSelection.tex input file
\subsection{Object reconstruction}
\label{sec:objreco}
While the four different channels use different combinations of objects, the object reconstruction is standardized across all of the different channels.
Three different objects are used: jets, $b$-tagged jets, and photons.

Selected events are required to contain a primary vertex with at least two associated tracks with transverse momentum $\pt>0.5~\GeV$~\cite{ATL-PHYS-PUB-2015-026}. The vertex with the highest sum of $\pt^2$ of the associated tracks is taken as the primary vertex.

Jets in this analysis are reconstructed from particle flow objects~\cite{PERF-2015-09} using the \akt algorithm~\cite{Cacciari:2008gp} as implemented in \Fastjet~\cite{Fastjet}, using a jet radius parameter $R=0.4$.
The ATLAS particle flow algorithm combines measurements from the ATLAS inner detector and calorimeter systems~\cite{PERF-2014-07}
to improve the jet energy resolution, reduce sensitivity to \pileup effects, and improve jet reconstruction efficiency, especially at low jet \pt.
The jet energy scale of particle flow jets is calibrated with a combination of simulation-based and \emph{in situ} corrections~\cite{JETM-2018-05}.
Calibrated jets are required to have a $\pt > 25$~\GeV~and satisfy $|\eta|<2.5$.
To reduce the effects of \pileup, jets with $\pt < 60$~\GeV and $|\eta| < 2.4$ are required to satisfy the ``tight'' working point of the jet vertex tagger criteria~\cite{PERF-2014-03},
which selects jets originating from the selected primary vertex.
A set of quality criteria is also applied to reject events containing at least one jet arising from non-collision sources or detector noise~\cite{ATLAS-CONF-2015-029}.

Reconstructed jets are $b$-tagged using the DL1r multivariate algorithm~\cite{FTAG-2019-07}.
Jets are considered $b$-tagged if they satisfy the 77\% $b$-tag efficiency working point, as measured in inclusive $t\bar{t}$ events.
The corresponding rejection factors for gluon/light-quark jets and charm-quark jets are approximately 200 and 5, respectively.
This working point was found to result in optimal signal sensitivity for both the \abb~and \jbb~channels.

Photons are reconstructed from energy deposits in the electromagnetic calorimeter~\cite{EGAM-2018-01}.
Reconstructed photon candidates are required to have $\pt > 25$~\GeV, and only central photons with $|\eta|<2.5$ are considered, excluding the barrel/endcap calorimeter transition region ($1.37<|\eta|<1.52$).
Only photons that satisfy the \emph{tight} identification requirement are considered, and additional tight isolation requirements are imposed to further suppress contamination from jets~\cite{EGAM-2018-01}.

An overlap removal procedure is applied to avoid any double-counting between the reconstructed photons and jets, by removing any jet with an angular distance $\sqrt{\Delta \eta^2 + \Delta \phi^2}<0.4$ to any photon.

\subsection{Photon channel selection}

The two photon channels (\ajj~and \abb) are required to satisfy the singel-photon trigger~\cite{TRIG-2018-05} with the lowest threshold where all events passing it are saved (unprescaled), and a single photon with $\pt > 150$~\GeV~is required in order for the trigger selection to be fully efficient.
In addition, the event is required to have two reconstructed jets passing the selection in Section~\ref{sec:objreco}, and the two highest-\pt (leading) jets are used to reconstruct the resonance candidate.
Several dijet and $\gamma$-jet kinematic variables were studied to enhance the $\Zp\rightarrow q\bar q$ signal over the expected background, and the strongest discriminating variable was found to be the asymmetry $\asymm=\frac{\left| y_1 - y_2\right|}{2}$, where $y_1$ and $y_2$ are the rapidities of the two selected jets.
It is expected that a signal from \Zp production produces isotropic jets while forward jets are more common in background events, causing the signal to have smaller values of \asymm~ than the background.
The shape of the $\asymm$ variable is found to be well described by the background simulation, with maximum differences of a few percent.
A cut of $\asymm<0.825$ is applied to maximize the signal sensitivity across the different mass ranges.
For the \abb~channel, in addition to satisfying the selection for the \ajj~channel, the two leading jets in the event are required to satisfy the $b$-tagging requirements described in Section~\ref{sec:objreco}.

\subsection{Trijet channel selection}

In both the \jjj~and \jbb~channels, events are required to have at least three reconstructed jets, where the leading jet is required to be above 475~\GeV  to be fully efficient for the lowest unprescaled single jet trigger~\cite{TRIG-2012-01}.
When considering only the three leading-\pt jets in a given event, there are three possible choices to assign two jets as corresponding to the resonance decay product candidates.\footnote{While a fourth jet can contribute, it is typically not the dominant contribution, and so it is not considered, to simplify the combinatorics.}
In the case where the resonance decays into two $b$-quarks, the combinatorial problem can be in principle trivially solved by identifying the two $b$-tagged-jets in the event.
While possible, this causes challenges in background modeling, and so this trivial solution is not used.
To illustrate the combinatorial problem more clearly, Figure~\ref{combinatorics_contribution} shows the particle-level dijet mass spectrum of different jet pairs for three different \Zp~signal samples.
Each jet is assigned a number based on its \pt ordering, e.g., $m_{12}$ is the dijet mass distribution (\mjj) of the leading and subleading jets.
For lower \Zp~mass, the $m_{23}$ contribution clearly peaks near the \Zp~mass, while at higher \Zp~mass, contributions from $m_{13}$ become more important.

\begin{figure}[tbp]
\centering
\subfigure[]{\includegraphics[width=0.45\textwidth]{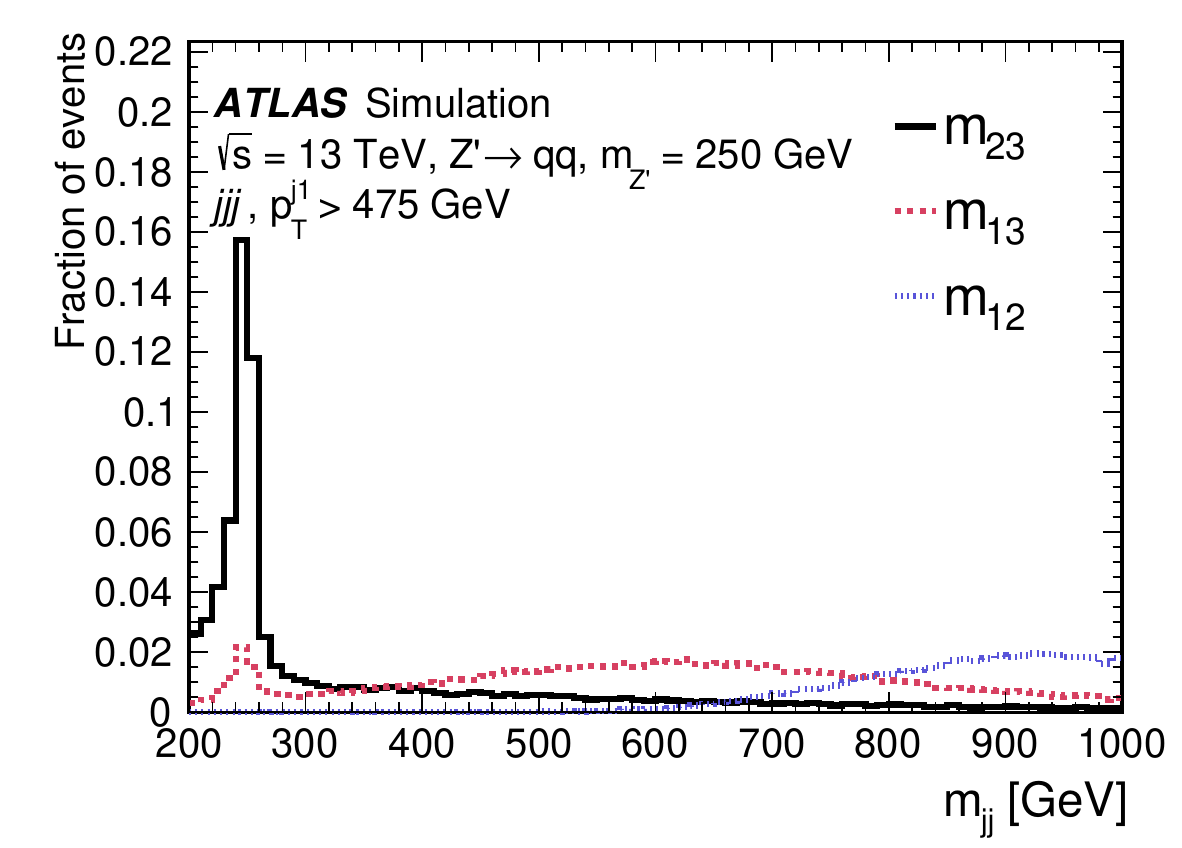}}
\subfigure[]{\includegraphics[width=0.45\textwidth]{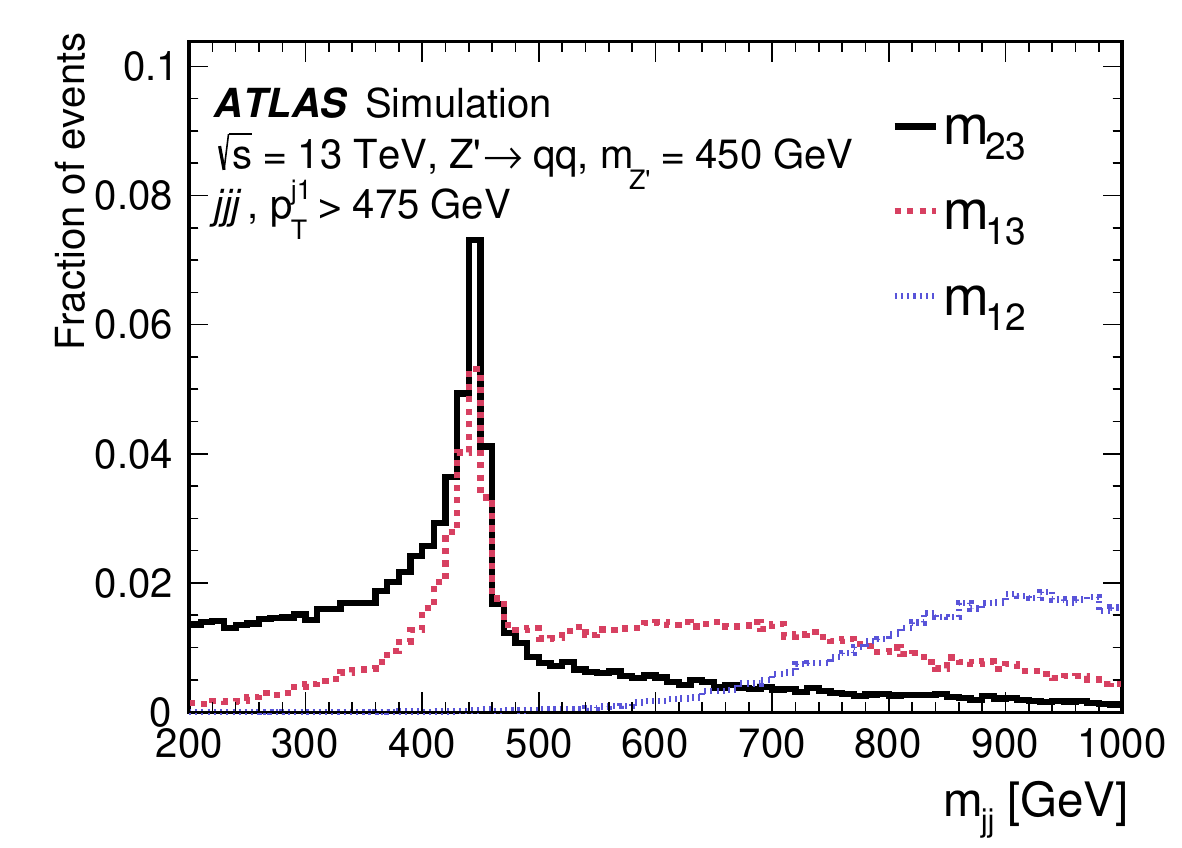}}
\subfigure[]{\includegraphics[width=0.45\textwidth]{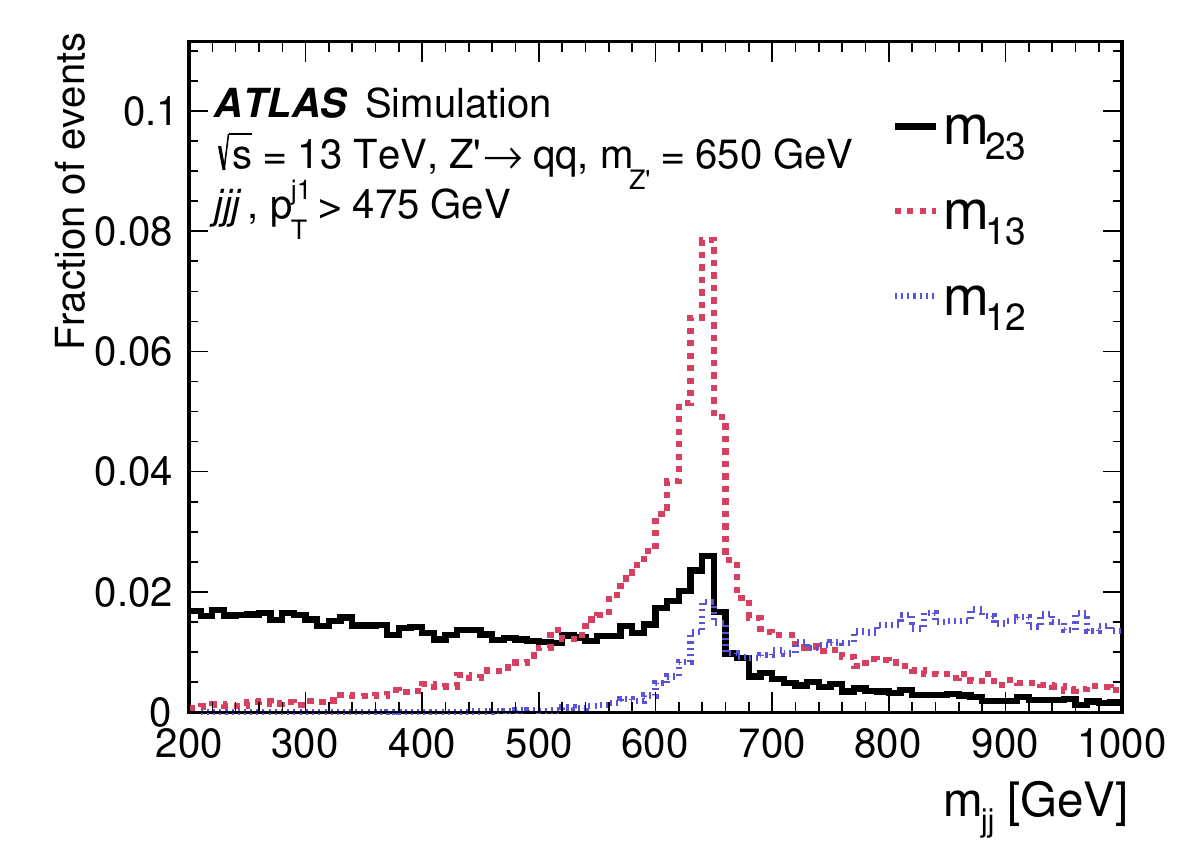}}
\caption{Distributions of particle-level dijet mass in simulated $\Zp \rightarrow qq$ events corresponding to different combinations of two jets, where jets 1,2 and 3 correspond to the leading, sub-leading and third-leading jets. Signal events generated with (a) $m_{\Zp}=250~\GeV$, (b) $m_{\Zp}=450~\GeV$ and (c) $m_{\Zp}=650~\GeV$ are shown.} %
\label{combinatorics_contribution}
\end{figure}

The combinatorial challenge is further complicated by the background estimate, which relies on a functional fit to a smoothly falling \mjj spectrum,
meaning that any additional selection to solve the combinatorial problem must also result in a smooth \mjj~spectrum.
The dominant background for these channels may be simplistically modeled as balanced high-\pt dijet events where one jet emits a softer jet, resulting in a trijet event.
Since the emitting jet loses some energy to its emission, the three jets are roughly \pt ordered from low to high as the emitted jet, the emitting jet, and the non-emitting jet,
it is possible to roughly map $m_{23}$ to $m_\text{emitted,emitting}$, $m_{13}$ to $m_\text{non-emitting,emitted}$ and $m_{12}$ to $m_\text{non-emitting,emitting}$.
In reality, this description of the background does not fully describe the observed background.
Multiple emissions may occur, and various detector effects can spoil the \pt balance of the reconstructed jets, meaning that the assumption that $m_{23}$ corresponds to the mass of the emitting and emitted jet may not hold in some events.

The mass spectra of the three possible pairings have different shapes, since different underlying physics contributes to each of them.
This is illustrated in Figure~\ref{fig:trijet:mjjComparison:all}, which shows the \mjj~distribution for $m_{12}$, $m_{13}$, and $m_{23}$ distributions.
Any generic combination of the different pairings will result in non-smoothness in the spectrum, roughly corresponding to the turn-ons in these different pairings.
Hence instead of pairing jets based solely on \pt-ordering, a different pairing method based on the kinematics of the three leading jets need to be used in the analysis.

Three similar options for choosing the dijet pairing were compared to determine which one results in the smoothest spectrum, each roughly corresponding to the mass of the emitting and emitted jet:
the second- and third-leading jets (as described above), the jet pair with the minimum $|\Delta \phi|$, and the pair of jets with the lowest mass.
The smoothness of the \mjj~spectrum at low masses determines the fit range therefore limiting the lowest \Zp~mass reach covered by the search.
The signal efficiencies of these three \mjj~selection options are similar.
However, selecting jet pairs with the minimum $|\Delta \phi|$ between them results in the smoothest low-mass spectrum in studies of multijet MC simulation. %

One final aspect to consider is that in dijet-like topologies, where one of the high-\pt jets radiates a low-\pt gluon reconstructed as a jet,
the gluon radiation tends to happen at small angles, while the two high-\pt jets are produced back-to-back.
This means that the $|\Delta \phi|$ selection tends to identify the radiating jet and the gluon.
Depending on which jet radiates the gluon, this selection can identify either the second- and third-leading jets or the first- and third-leading jets.
In a small fraction of events, the $|\Delta \phi|$ selection identifies the two leading jets in the events as trijets, but where the underlying topological assumptions break down.
These events, comprising around 0.2\% of background events, do not produce a smooth \mjj~spectrum, as shown in Figure~\ref{fig:trijet:mjjComparison:m21}. %
Based on this, events where the $|\Delta \phi|$ selection identifies the two leading jets are  rejected to avoid biasing the fit.

\begin{figure}[tbp]
\centering
\subfigure[]{\label{fig:trijet:mjjComparison:all}\includegraphics[width=0.45\textwidth]{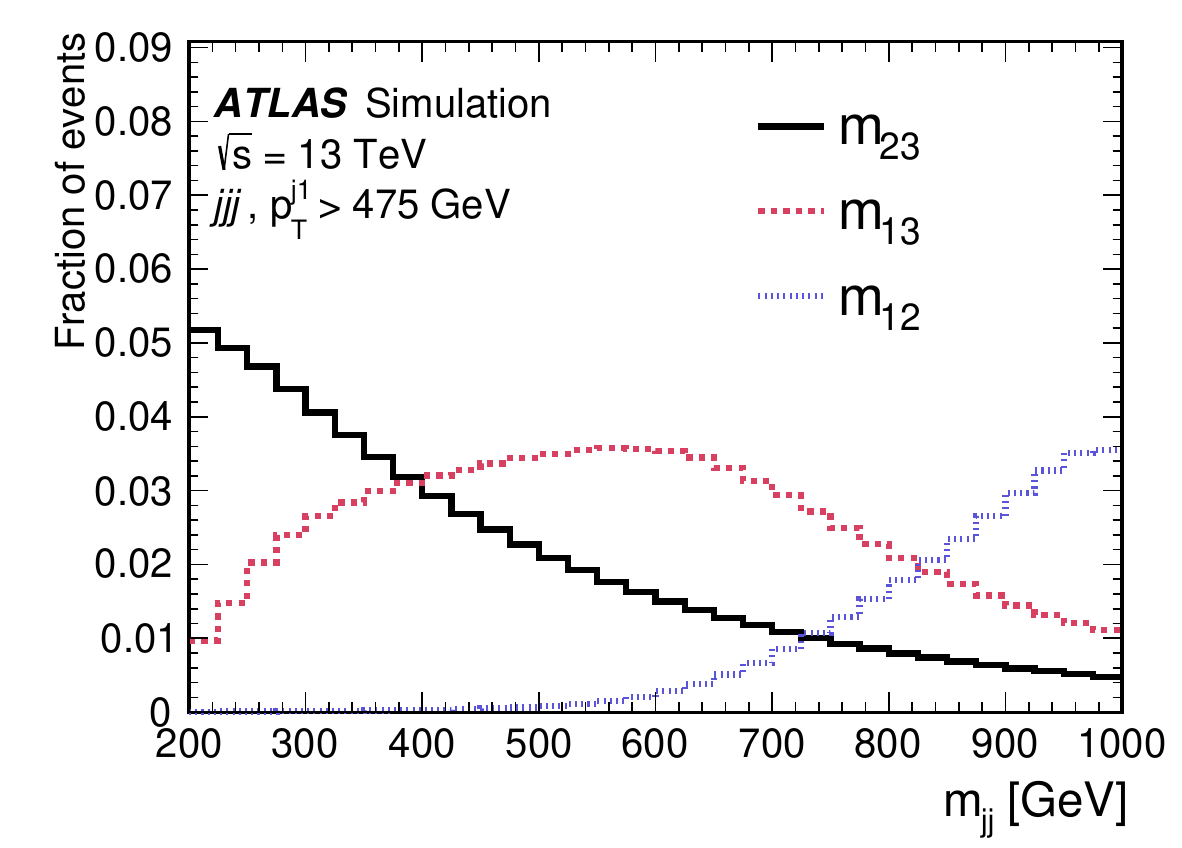}}
\subfigure[]{\label{fig:trijet:mjjComparison:m21}\includegraphics[width=0.45\textwidth]{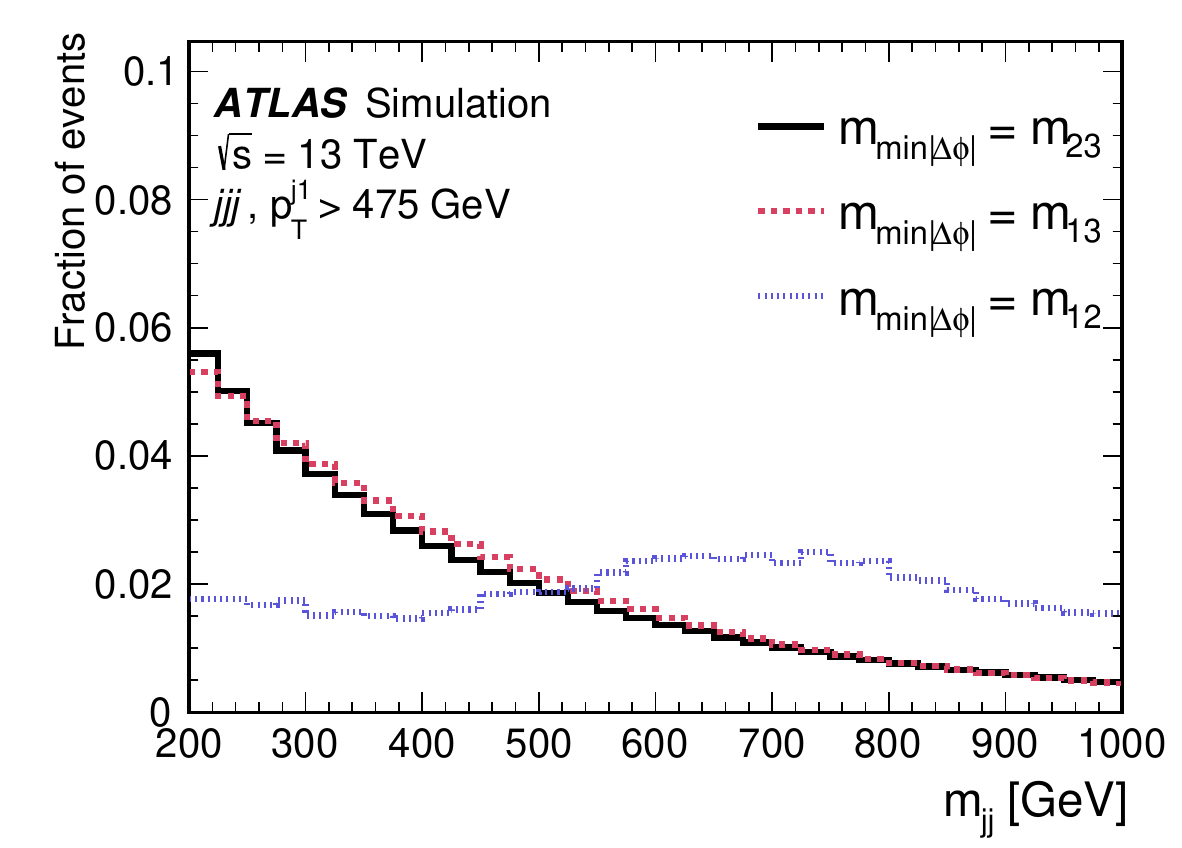}}
\caption{(a) Distributions of reconstructed dijet mass in simulated background events corresponding to different combinations of two jets, where jets 1,2 and 3 correspond to the leading, sub-leading and third-leading jets, and (b) comparison of the shape of the normalized selected dijet \mjj~spectra in the case where the selected jets are the second and third leading jets ($m_{\text{min}|\Delta \phi|}=m_{23}$), the first and third leading jets ($m_{\text{min}|\Delta \phi|}=m_{13}$), or the first and second leading jets ($m_{\text{min}|\Delta \phi|}=m_{12}$). Events are simulated with~\PYTHIA.}
\end{figure}

To summarize, the algorithm selected to identify the \Zp~decay candidates
takes the jet pair with the minimum $|\Delta \phi|$ between them and rejects the event if such jets correspond to the two \pT-leading jets.
The pair of jets passing such requirements are referred to as the \textit{selected jets} and, as explained above, produces the smoothest low-mass spectrum in studies of multijet MC simulation.
No additional background suppression is applied, since cuts on observables such as \asymm~can sculpt the background distribution.
For the \jbb~channel, in addition to passing the same jet requirements as the \jjj~channel, the two jets with the minimum $|\Delta \phi|$ between them are also required to be $b$-tagged. %

The resulting signal \mjj~distributions after the event selection are shown for each channel in Figure~\ref{fig:signalShapes}.
The \jjj~channel, at higher masses, has a clear tail in the \mjj~distribution resulting from the incorrect selection of the dijet pair.
This tail is less pronounced for the \jbb~channel, owing to the additional $b$-tagging requirement imposed on the selected jets.
The signal efficiency for the \jbb~selection gets smaller at high masses, as the selected dijet pair is often not $b$-tagged there. Such an effect is considered less relevant than obtaining a smoothly falling \mjj~distribution for the background at lower \mjj~values, as the more energetic part of the \mjj~spectrum is better studied by other analyses~\cite{EXOT-2016-20,CMS-EXO-14-005}.

\begin{figure}[tbp]
\centering
\subfigure[]{\includegraphics[width=0.45\textwidth]{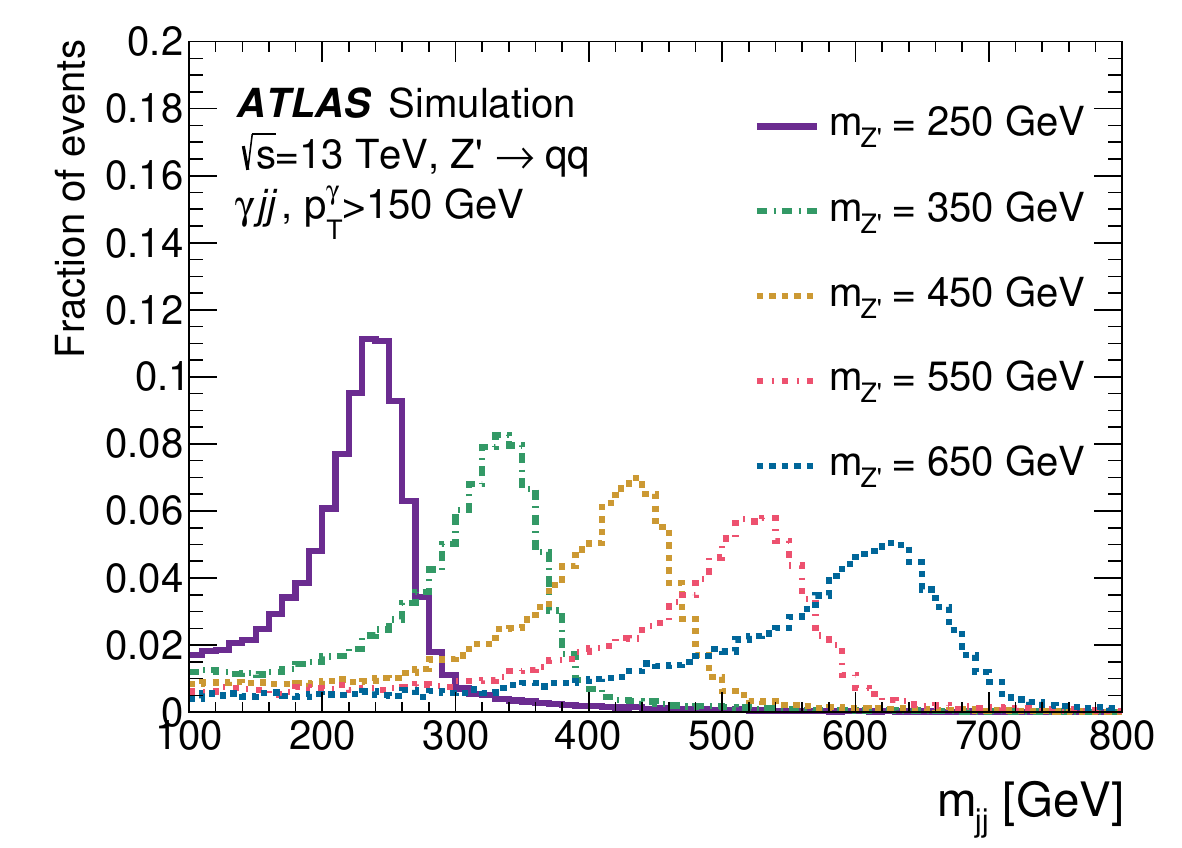}}
\subfigure[]{\includegraphics[width=0.45\textwidth]{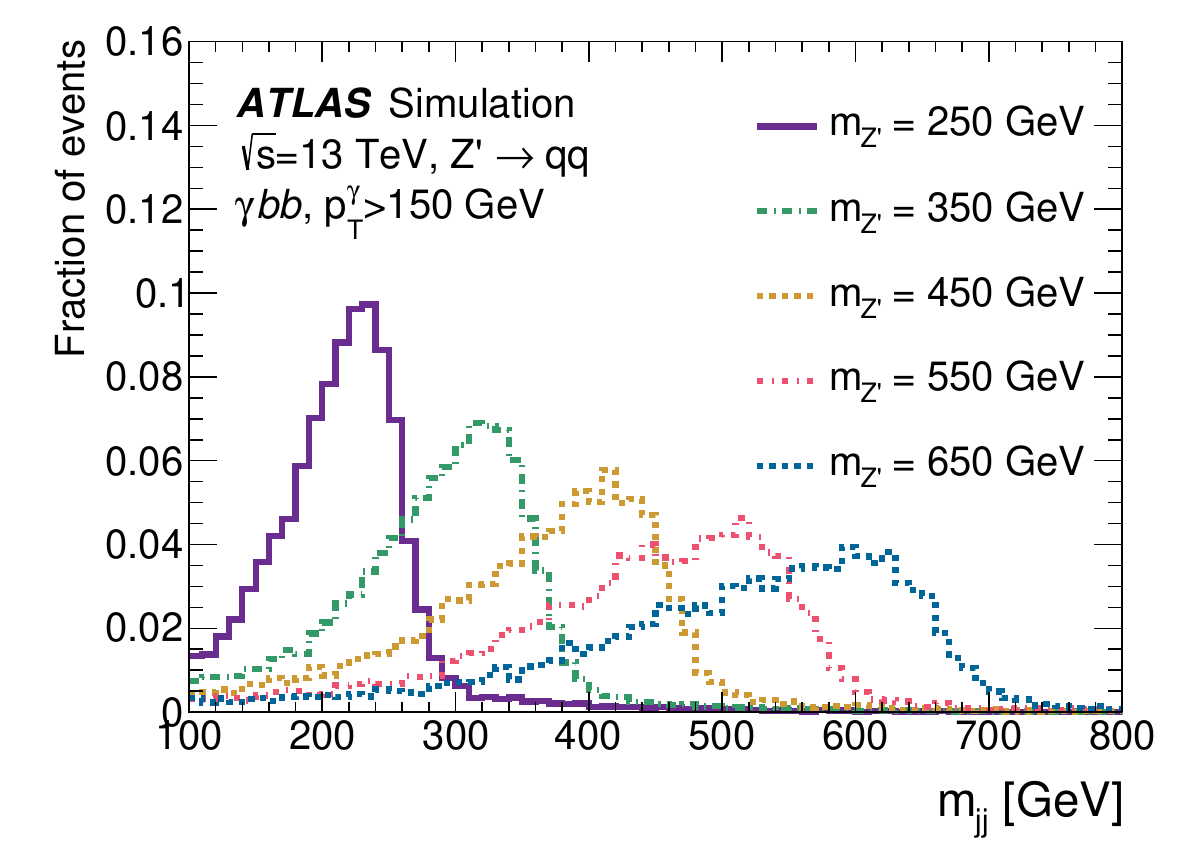}}
\subfigure[]{\includegraphics[width=0.45\textwidth]{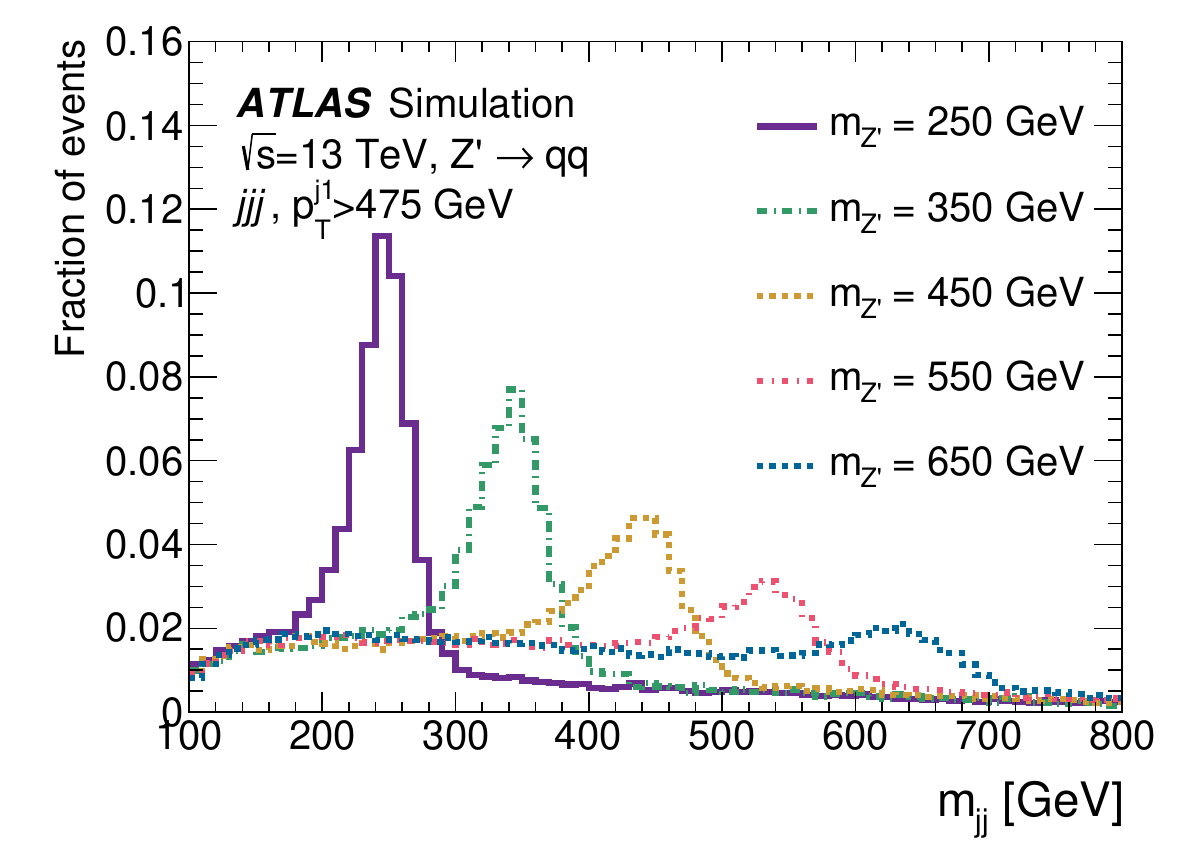}}
\subfigure[]{\includegraphics[width=0.45\textwidth]{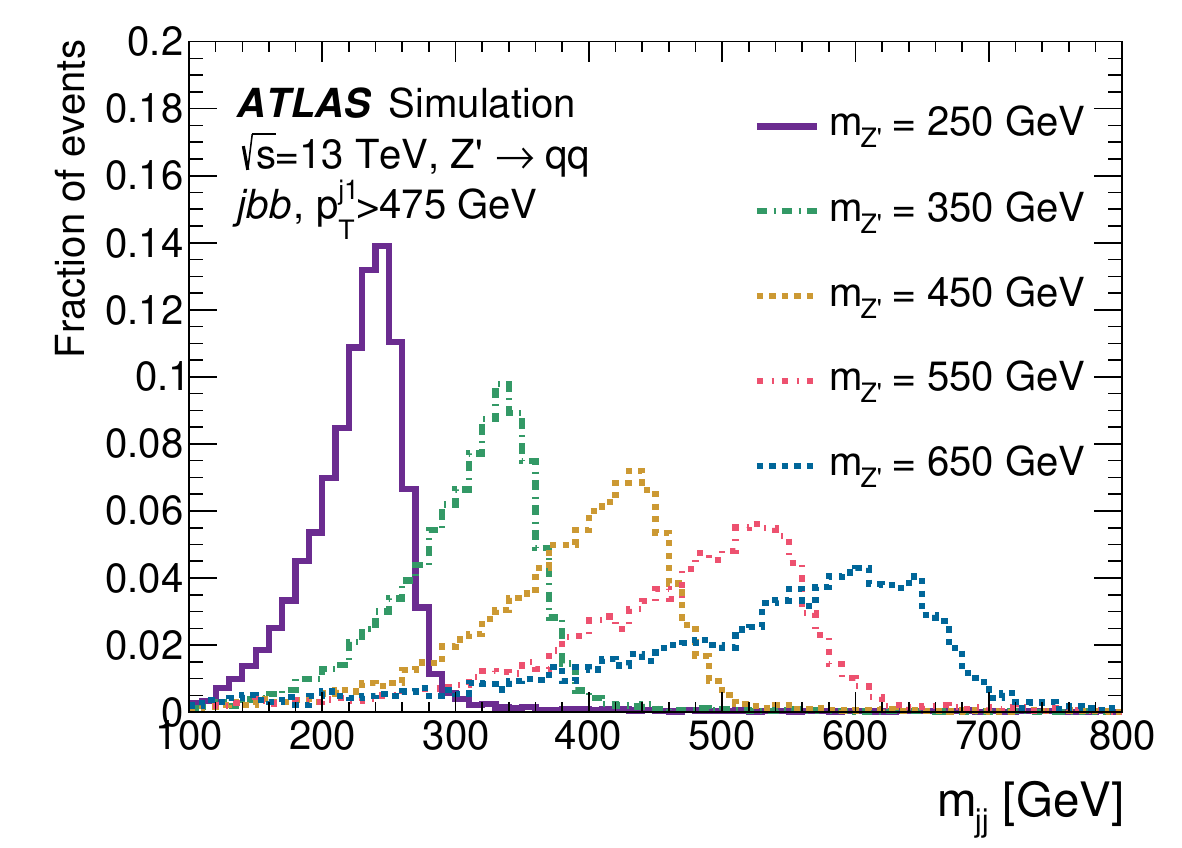}}
\caption{Distributions of reconstructed dijet mass in simulated $\Zp\rightarrow qq$ events with different $m_{\Zp}$ values in the (a) \ajj, (b) \abb, (c) \jjj, and (d) \jbb~channels after full event selection.}
\label{fig:signalShapes}
\end{figure}

% End of text imported from the .//sections/eventSelection.tex input file

\FloatBarrier
\section{Methodology}
\label{sec:methodology}

% The next lines are included from the .//sections/method.tex input file

The numbers of signal and background events are estimated channel-by-channel by binned profile likelihood fits of a signal-plus-background model to the measured \mjj~distributions.
The number of selected events as a function of \mjj~is parameterized  in the fit as:
\begin{equation*}
n_\text{tot} (\mjj)=n_{S} \times f_{S} (\mjj) + n_{B} \times f_{B} (\mjj) \quad ,
\end{equation*}
where $n_{S}$ and $n_{B}$ are the number of signal and background events, respectively, and $f_{S}(\mjj)$ and $f_{B}(\mjj)$ are their probability distribution functions. All the distributions entering the fit are binned finely in bins of 1~\GeV~width, to minimize the loss of information and the typical biases of binned likelihood fits~\cite{DEMBINSKI2019135}.

\subsection{Signal templates}

The likelihood fit results are interpreted using model-independent and model-dependent strategies.
For the model-independent limits, the signal probability distribution functions are taken as Gaussian distributions, with a mean equal to the resonance mass, with widths ranging from 5\% to 15\%.
The upper end of the template width is determined based on the results of the spurious signal test and signal injection test described in Section~\ref{sec:method:background}.

For the model-dependent limits, the shape of the~\mjj~distributions from simulated samples are used directly.
Signal samples are produced with a limited set of signal masses, and these templates are used as inputs to an interpolation between mass points to provide a finer signal grid
using the method described in Ref.~\cite{Baak:2014fta}.

\subsection{Background estimate}\label{sec:method:background}

The selected jets in non-resonant QCD processes, which constitute the SM background for this search, result in dijet systems with smoothly-falling invariant mass distributions.
In order to estimate this background in the search regions, a class of parametric functions is fit to the \mjj~distributions:
\begin{equation*}
f_{B}(x) = p_{1} (1-x)^{p_{2}} x^{p_{3} + p_{4} \ln(x) + p_{5} \ln^2(x) + p_{6} \ln^3(x)},
\end{equation*}
where $x=m_{jj}/\sqrt{s}$, and $p_{1}, p_{2}, p_{3}$, $p_{4}$, $p_{5}$, and $p_{6}$ are the fitted parameters.
For an $N$-parameter fit, only the first $N$ parameters are allowed to vary, while the higher-$N$ terms are fixed to zero.
Such functions were successfully used by a wide variety of dijet and multijet resonance searches by the
CDF, CMS, and ATLAS Collaborations~\cite{CDF:2008ieg,EXOT-2010-01,CMS-EXO-11-015,EXOT-2013-11,EXOT-2015-02,CMS-EXO-16-032,EXOT-2020-15,EXOT-2019-03,CMS-EXO-21-010,EXOT-2022-18}.

The data-driven background fitting procedure employs a 5-parameter function and was validated using several cross-checks, including spurious signal tests and signal injection tests.
These tests were performed using simulated samples, with a subset of checks performed on a partial data sample.

The spurious signal test evaluates whether the fitting procedure is biased such that it produces a non-zero extracted signal when fitting a data sample with no true signal.
This test is performed using many pseudodata distributions which are generated from background-only fits to the simulated \mjj~distribution with a $6$-parameter function,
which has more flexibility than the final fit function.
Each distribution is tested by performing signal+background fits with  a $5$-parameter fit function and various signal hypotheses.
Gaussian-shaped signals with widths ranging from 5 to 15\% of the signal mean are tested, as well as using the signal templates from the \Zp~simulated samples directly.
For each pseudodata distribution and signal hypothesis $n_{S}$ is determined, and the median value and standard deviation of the $n_{S}$ distribution
are taken to be $S_{\mathrm{spur}}$ and $\sigma_\mathrm{spur}$ respectively.
To satisfy the spurious signal requirements, $S_\mathrm{spur} / \sigma_\mathrm{spur}$ is required to be less than 0.5 for every signal hypothesis mass point.
This requirement is satisfied for all channels for \Zp~masses between 250~\GeV and 650~\GeV.
While the fits extend to \mjj~values beyond 650~\GeV, higher signal masses do not always satisfy the spurious signal tests, and are not considered further.
In addition, for the \jjj~channel, the signal shape degrades significantly at higher signal masses, since the selected jets often do not correspond to the resonance decay products.
Finally, the \jbb~channel passes this requirement for resonance masses down to 200~\GeV, which is possible since the \mjj~spectrum in this channel is smooth down to lower masses than the other channels.

The signal injection test is performed to ensure that the fit is able to extract a signal component with the expected signal strength.
Simulated signal models, with both Gaussian shapes and in a range of widths and signal templates, are included in the fitted background distribution with a given signal cross-section selected to be in the range $0-5\sigma$,
where $\sigma=n_{S}/\sqrt{n_{B}}$ and the number of signal and background events are determined using a $2\sigma$ window around the injected signal peak in each test.
The injected signals in this study were extracted for pseudodata distributions, with the requirement that the median of the extracted significances
is within $0.5\sigma$ of the injected significance, for all the signal hypotheses and mass points individually.
These tests were satisfied for all channels for the same range of resonance masses and widths as with the spurious signal tests, while they failed for Gaussian-shaped signals with widths above 15\%.

The fitted \mjj~region is determined separately for each channel.
For the chosen range, the fit quality, spurious signal, and signal injection tests are must be satisfied for the MC simulation.
The fitted \mjj~region is chosen to prioritize a fit that starts as low in \mjj~as possible while maintaining a wide enough fit range to study the relevant signal models.
The lowest part of the \mjj~spectrum does not have a smoothly falling shape because of trigger inefficiency and analysis selection effects, such as the $\asymm$ cut, and such a behavior cannot be fit by the class of functions considered.
The \ajj~and \abb~channels are found to have a smoohly falling behavior starting from $\mjj=200~\GeV$, while the \jjj~and \jbb~channels from $\mjj= 225~\GeV$ and $\mjj= 160~\GeV$ respectively.
The upper fit ranges are defined from the edges of bins defined by the \mjj~resolution, and are taken to be the highest among those where the background fit functions to the simulated \mjj~distributions have a $\chi^2 p$-value above 0.05. The procedure to determine the fit range is repeated on the measured data, to ensure the fit ranges estimated from the MC simulation are adequate. If they are not, the upper limit of the fit range is reduced by one \mjj~resolution bin, and all the tests are repeated.

For the \abb~channel, the background estimate is further validated using an ABCD method with simulated samples, following the strategy of Ref.~\cite{EXOT-2018-05}.
Such method allows a test of the background estimate in the \abb~channel with a higher statistical precision than available in the existing simulated sample and also serves as a test for potential biases in the \mjj~spectrum introduced by $b$-tagging.
The ABCD re-weighting method defines different regions based on whether events satisfy or fail to meet the \asymm~selection and the $b$-tagging requirements, providing an estimate for the $b$-tagged spectrum which is based on the un-tagged distribution.
The ABCD-reweighted \mjj~ spectrum is found to be compatible with the one obtained by applying the \abb~event selection, within the statistical precision of the re-weighted and high-statistics spectrum. Given the level of agreement, no uncertainty related to $b$-tagging is considered to affect the background fit.
The resulting distribution is found to satisfy the spurious signal and signal injection tests described above, providing further confirmation of the fit strategy.
Since this method relies on the \asymm~selection, the technique cannot easily be applied to the \jbb~channel, but the conclusions on the possible $b$-tagging biases from \abb~apply to the \jbb~channel as well.

% End of text imported from the .//sections/method.tex input file

\section{Systematic  uncertainties} %
\label{sec:systematics}

% The next lines are included from the .//sections/systematics.tex input file

When interpreting the analysis in terms of candidate signal models, the impact of various experimental and theoretical sources of uncertainty is considered.

\textbf{Luminosity.}
The uncertainty in the combined 2015--2018 integrated luminosity is 0.83\%~\cite{DAPR-2021-01}, obtained using the LUCID-2 detector~\cite{LUCID2} for the primary luminosity measurements.
It is treated as a single normalization uncertainty applied as a scale factor to the signal models.

\textbf{Jets.}
Systematic uncertainties in the $R=0.4$ jet energy scale and resolution are evaluated using a series of \emph{in situ} measurements and simulation-based techniques, thoroughly documented in Ref.~\cite{JETM-2018-05}.
The most significant source of uncertainty in the JES originates from uncertainties on the flavor of the jet.
Uncertainties on the JVT are also included, but are negligible, since the uncertainties are applied only to the signal.
For the photon channels, these uncertainties have a negligible impact on the efficiency, and around a 2\% impact on the location of the \Zp~mass.
For the trijet channels, the impact on the location of the \Zp~mass is around 1\%, while the impact on the selection efficiency is around 3\%.

\textbf{Photons.}
The systematic uncertainties in the photon identification and isolation efficiencies are estimated following the prescriptions in Ref.~\cite{EGAM-2018-01}.
They are evaluated by varying the correction factors for photon selection efficiencies in simulation by the corresponding uncertainties and affect the diphoton selection efficiency.
The experimental uncertainties in the photon energy scale and resolution are obtained from Ref.~\cite{EGAM-2018-01}, and are only applied to the \ajj~and \abb~channels.
These systematic uncertainties primarily impact the selection efficiency of the signal events, with a total impact on the efficiency of around 2\%.

\textbf{B-tagging.}
For final states with requirements on the number of $b$-tagged jets, additional systematic uncertainties are applied.
The systematic uncertainty in the $b$-tagging efficiency is measured using data enriched in \ttbar events for jet $\pt< 400$~\GeV and extrapolated to higher \pt regions using a method similar to the one
described in Ref.~\cite{ATL-PHYS-PUB-2021-003}.
The impact of this systematic uncertainty in the efficiency of the event selection of signal events ranges from 2--3\%.

\textbf{Parton distribution functions.}
The theoretical uncertainty envelope associated with the \NNPDF[2.3lo] PDF set is propagated through the analysis, where their impact is primarily on the  normalisation of the signal events.
The change in analysis selection efficiency is recalculated for each provided PDF variation, and the standard deviation of all such variations is taken as a measure of the systematic uncertainty due to PDFs.
This uncertainty affects signal samples and ranges from 1--5\% across the different channels, with larger effects on the photon channels than on the dijet channels.

\paragraph{Background modeling}

A systematic uncertainty to cover potential biases in modeling the background shape is accounted for using the spurious signal $S_\mathrm{modeling}$.
The value of $S_\mathrm{modeling}$ is determined as the envelope of $|S_\mathrm{spur}|$ over $\mjj$ and is considered to cover features of the $\mjj$ background spectrum which are not modeled by the background parametrisation chosen.
This is implemented in the likelihood fit as an additional signal contribution, such that
\begin{equation*}
N_\mathrm{signal}(m_{\Zp}) = \sigma \times A \times BR \times \mathcal{L} + S_\mathrm{modeling} (m_{\Zp}) \theta_\mathrm{modeling},
\end{equation*}
where $N_\mathrm{signal}(m_{\Zp})$ is the number of extracted signal events at a given $m_{Z'}$, $\sigma$, $A$, $BR$, and $\mathcal{L}$ are the signal cross-section, acceptance, signal branching ratio, and integrated luminosity, respectively;
and $\theta_\mathrm{modeling}$ is a nuisance parameter associated with the background modeling uncertainty.
The signal acceptance is defined as the fraction of simulated events at detector level that pass the analysis selection cuts.

The spurious signal uncertainty is found to be, together with the statistical uncertainty, the leading source of uncertainty on the extracted number of signal events.

% End of text imported from the .//sections/systematics.tex input file

\section{Results}
\label{sec:results}

% The next lines are included from the .//sections/results.tex input file
\FloatBarrier

The dijet invariant mass distributions in data, together with the corresponding fitted background estimates, are shown in Figure~\ref{fig:bumphunter} for all four channels.
The data are well described by the 5-parameter fit function in all channels, and the global $\chi^2$ $p$-value ranges from 0.10 to 0.42 for the different channels.
Fit ranges are chosen such that the fits satisfy the tests explained in Section~\ref{sec:method:background}.
For the photon channels, the background fit spans the \mjj~range of 200~\GeV\ to 875~\GeV for the \ajj~channel and of 200~\GeV\ to 854~\GeV for the \abb~channel, while for the \jjj~and \jbb~channels, the fit spans 225~\GeV to 1000~\GeV and 160~\GeV~to 700~\GeV~respectively.

\begin{figure}[tbp]
\centering
\subfigure[]{\includegraphics[width=0.48\textwidth]{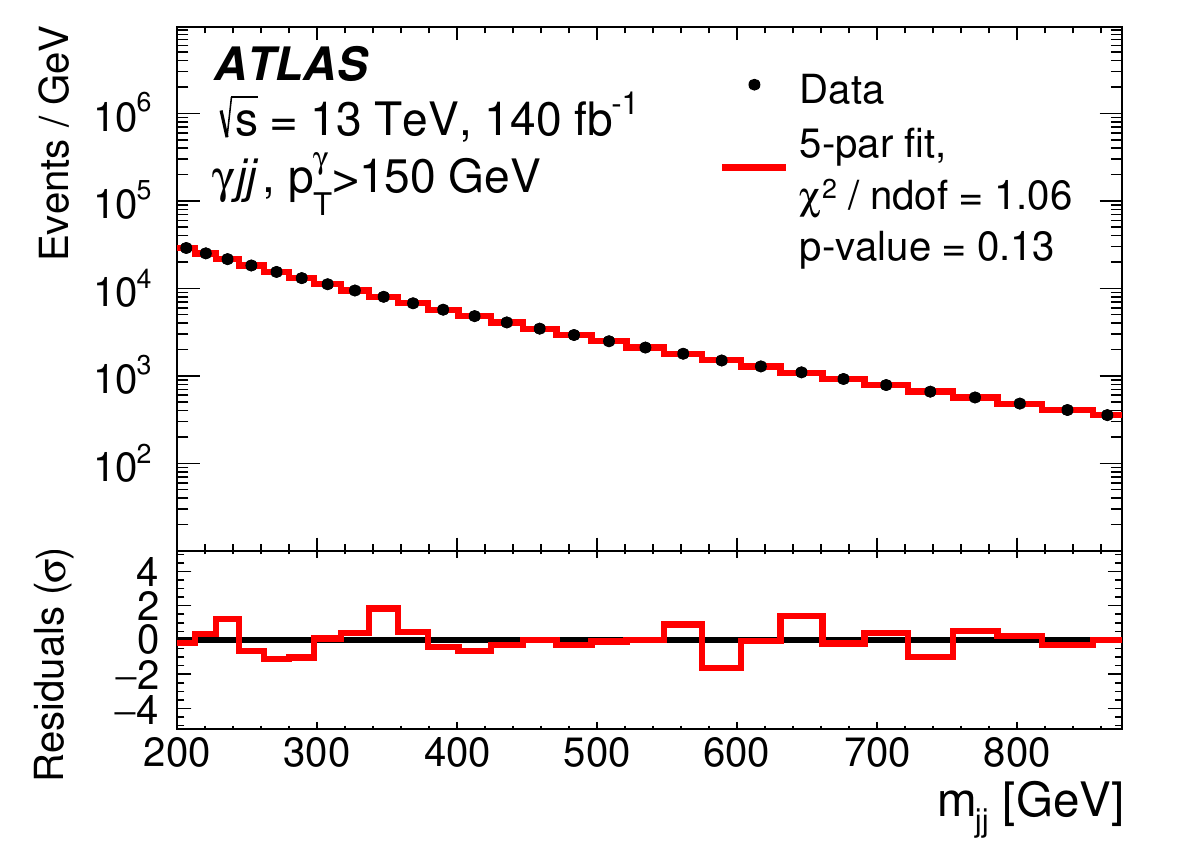}}
\subfigure[]{\includegraphics[width=0.48\textwidth]{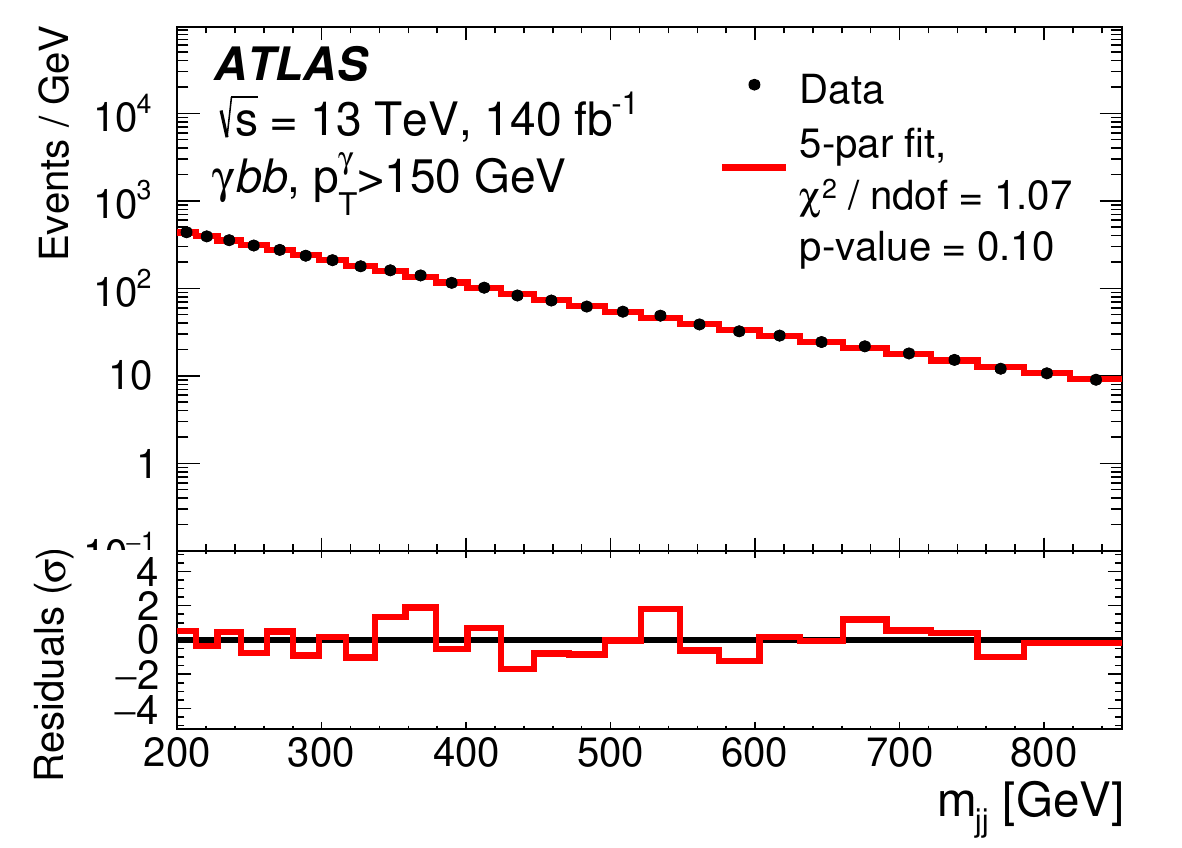}}
\subfigure[]{\includegraphics[width=0.48\textwidth]{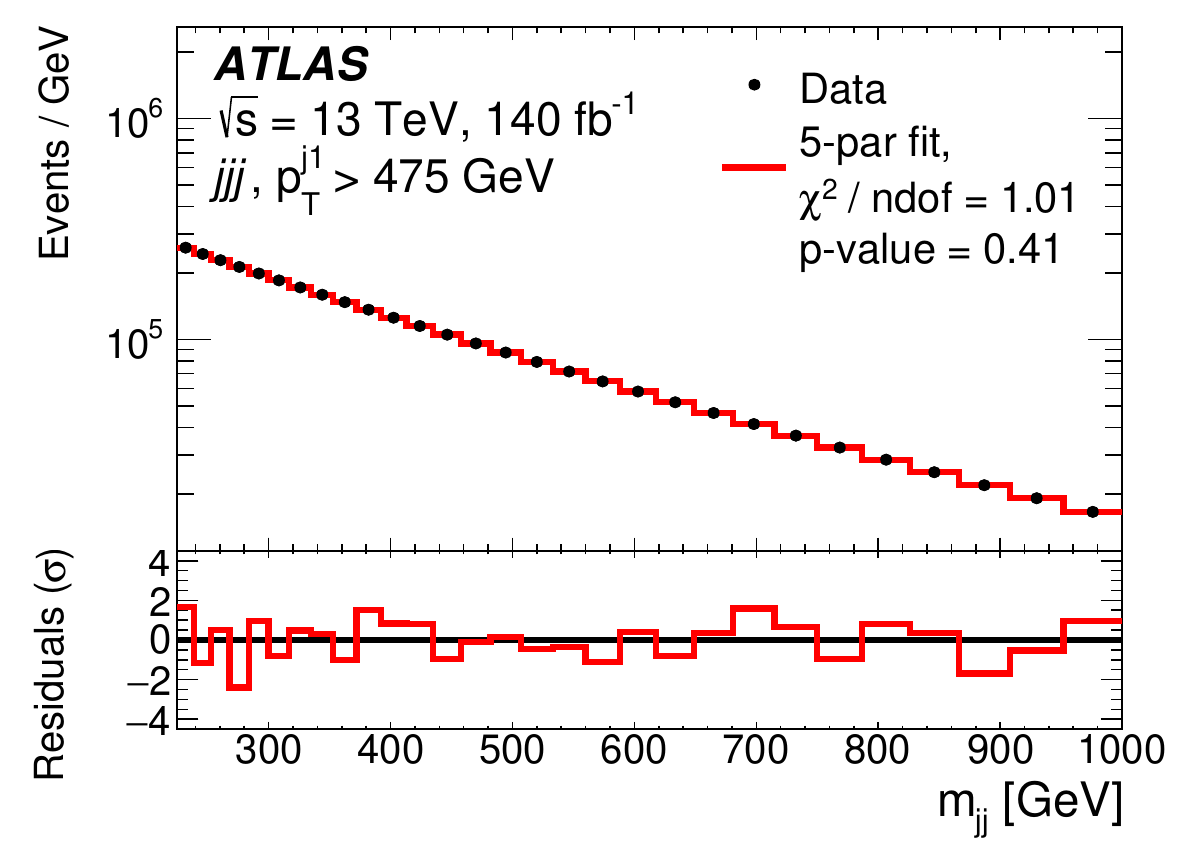}}
\subfigure[]{\includegraphics[width=0.48\textwidth]{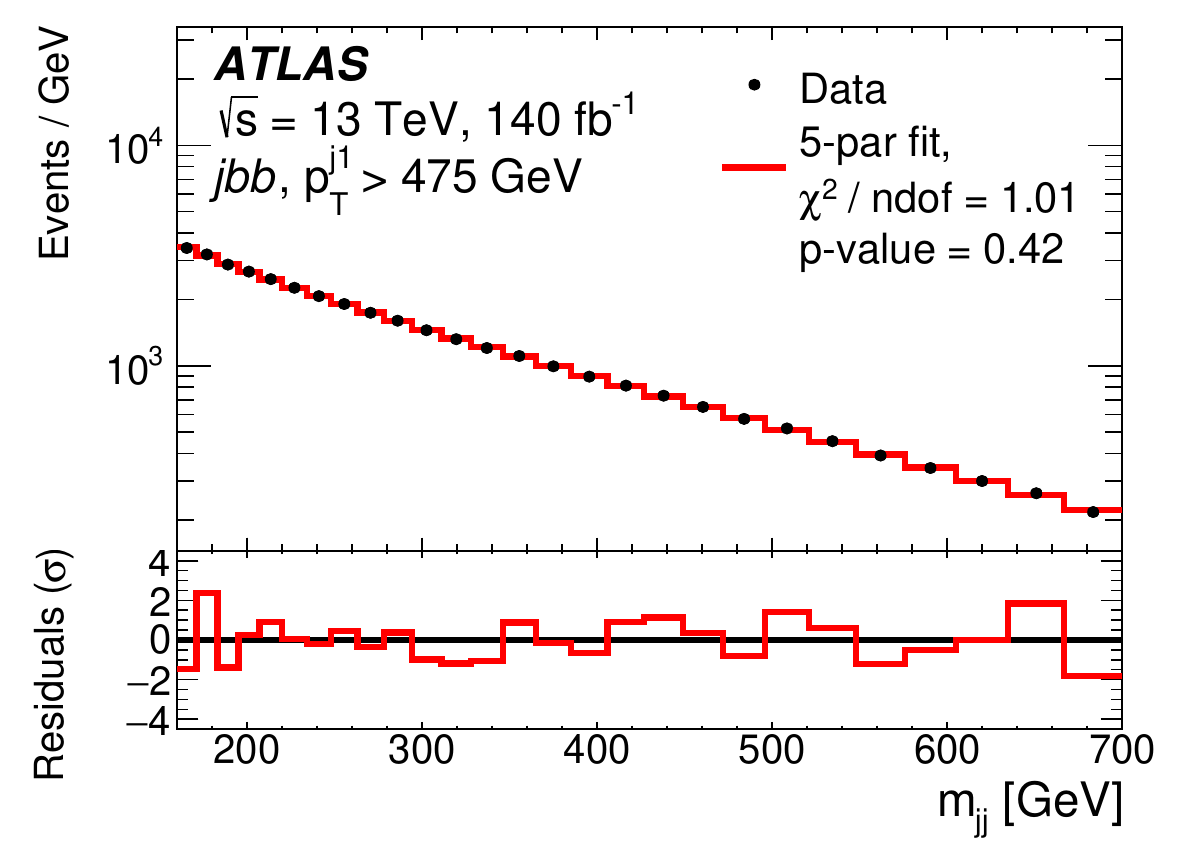}}
\caption{Dijet invariant mass distributions data compared to the fitted background estimates for the (a) \ajj, (b) \abb, (c) \jjj, and (d) \jbb~channels. Bottom panel shows the fit residuals in terms of standard deviations ($\sigma$). The distributions are shown here with the \mjj~resolution binning. }%
\label{fig:bumphunter}
\end{figure}

The \textsc{BumpHunter}~\cite{CDF:2008voc,Choudalakis:2011qn} algorithm, as implemented in \textsc{pyBumpHunter}~\cite{Vaslin:2022bds,pyBumpHunter},
is used to measure the statistical significance of localized excesses of the measured data relative to the estimated background in the \mjj~distributions, which could be due to the presence of resonant signals.
This is done using mass bins with a bin width determined by the mass resolution of ~\mjj~as a function of the mass, where the mass resolution is determined using a Gaussian function fit to the ~\mjj~response distribution.
Windows are allowed to have a width of up to three mass-resolution bins, hence corresponding to three times the \mjj~resolution, and for each scanned window
\textsc{BumpHunter} evaluates the statistical significance of the observed difference between the data distribution and the background fit.
The \textsc{BumpHunter} $p$-value is defined as the smallest observed probability for the data in a given window to deviate from the background prediction
by the observed amount due to a Poissonian fluctuation of the background, using pseudoexperiments generated from the background prediction, and without considering systematic uncertainties.
The most significant localized excesses identified by the \textsc{BumpHunter} algorithm are found at 348~\GeV with a local significance of 1.9$\sigma$ for the \ajj~channel, at 380~\GeV with a local significance of 1.8$\sigma$ for the \abb~channel, at 404~\GeV with a local significance of 1.8$\sigma$ for the \jjj~channel and at 522~\GeV with a local significance of 1.5$\sigma$ for the \jbb~channel.

As no significant deviation from the background expectation is observed, upper limits are set on the signal production rate as a function of the hypothesized resonance mass.
The numbers of signal and background events are estimated from maximum-likelihood fits of the signal-plus-background models to the corresponding \mjj~distributions.
Systematic uncertainties described in Section~\ref{sec:systematics} are included in the fits via nuisance parameters constrained by Gaussian-distributed penalty terms.
Systematic uncertainties affecting the signal shape have only a small impact on the result, and the dominant uncertainties arise from the statistical and spurious signal uncertainties.
The $p$-value is determined from a profile-likelihood-ratio-test statistic~\cite{Cowan:2010js}.
The local $p$-value for compatibility with the background-only hypothesis when testing a given signal hypothesis ($p_{0}$) is evaluated based on the asymptotic approximation~\cite{Cowan:2010js}.
Global significance values are computed from background-only pseudo-experiments to account for the trial factors due to scanning both the signal mass and the width hypotheses.
The expected and observed 95\% confidence level (CL) exclusion limits on the product of the cross-section, branching ratio, and acceptance are computed using a modified frequentist-approach~\cite{Read:2002hq},
in an asymptotic approximation to the test-statistic distribution~\cite{Cowan:2010js}.

Figure~\ref{fig:limits} shows the 95\% CL upper limits on the $\sigma \times A \times \mathrm{BR} $ of the \Zp~axial-vector dark-matter mediator as a function of its mass, derived using the signal templates.
Results are interpolated linearly in the log of the cross-section.
Similar results are shown in Figure~\ref{fig:limits:gaus} for generic signals with Gaussian function shapes with 5\%, 7\%, 10\%, 12\%, and 15\% signal widths.

\begin{figure}[tbp]
\centering
\subfigure[]{\includegraphics[width=0.48\textwidth]{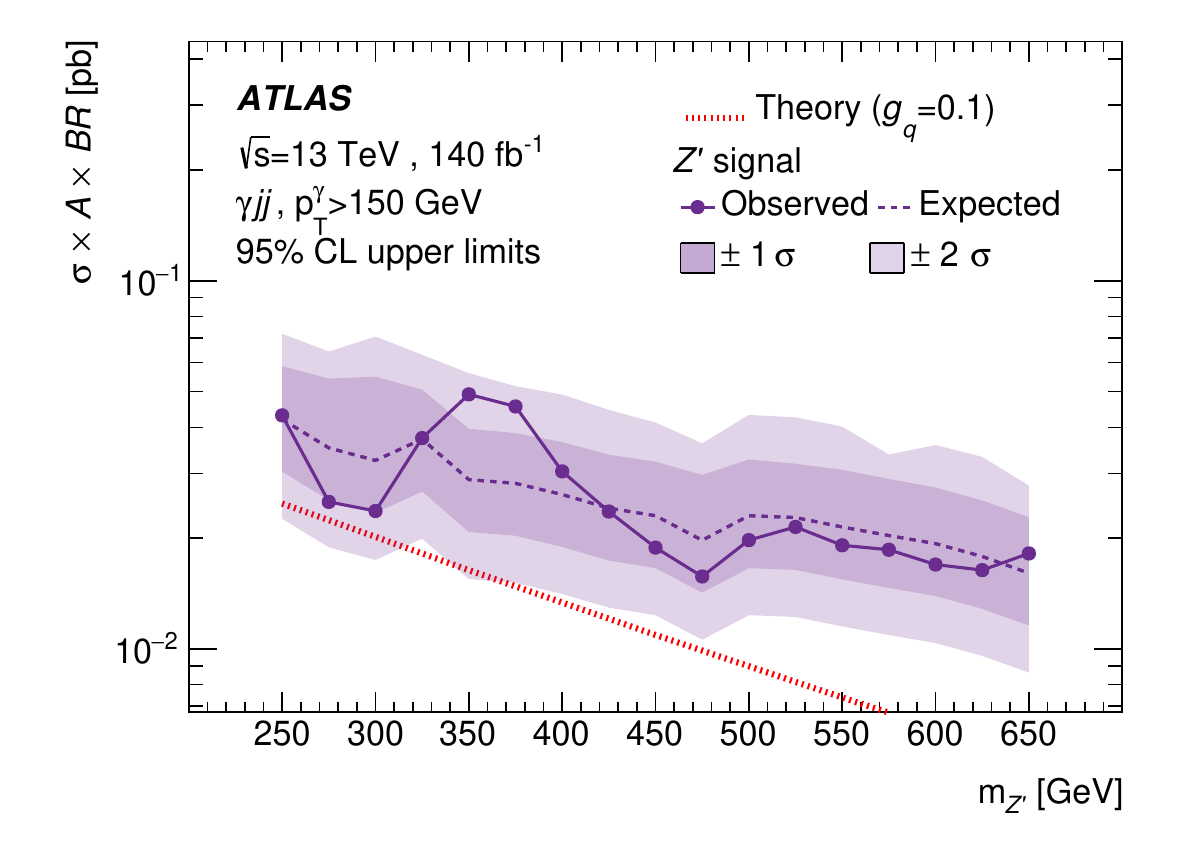}}
\subfigure[]{\includegraphics[width=0.48\textwidth]{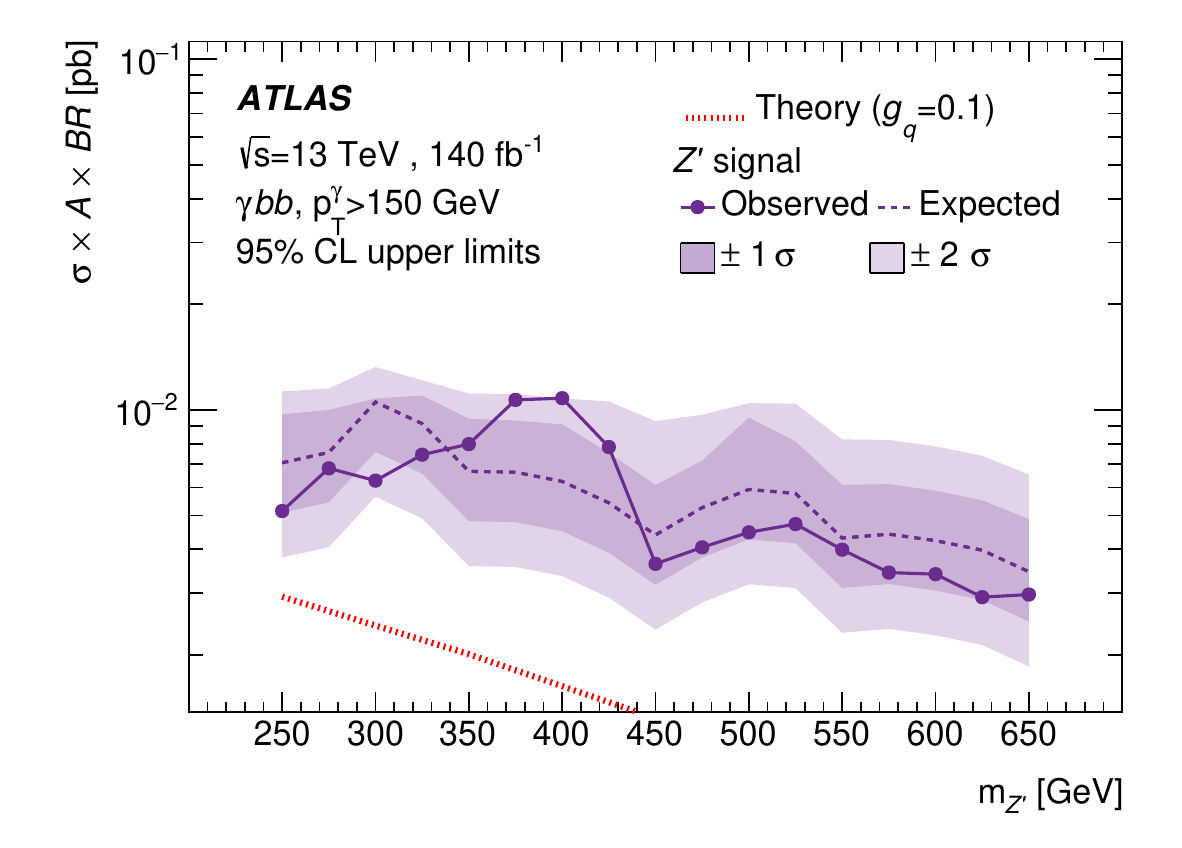}}
\subfigure[]{\includegraphics[width=0.48\textwidth]{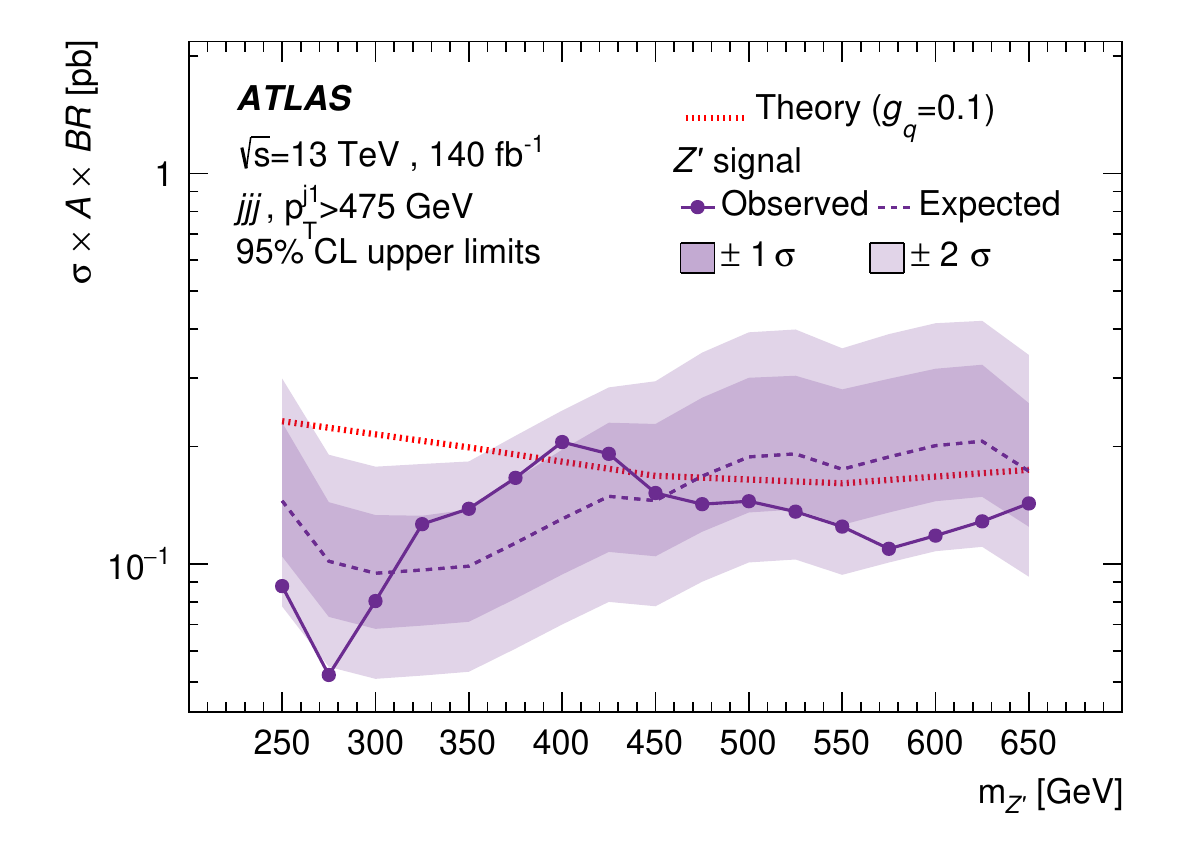}}
\subfigure[]{\includegraphics[width=0.48\textwidth]{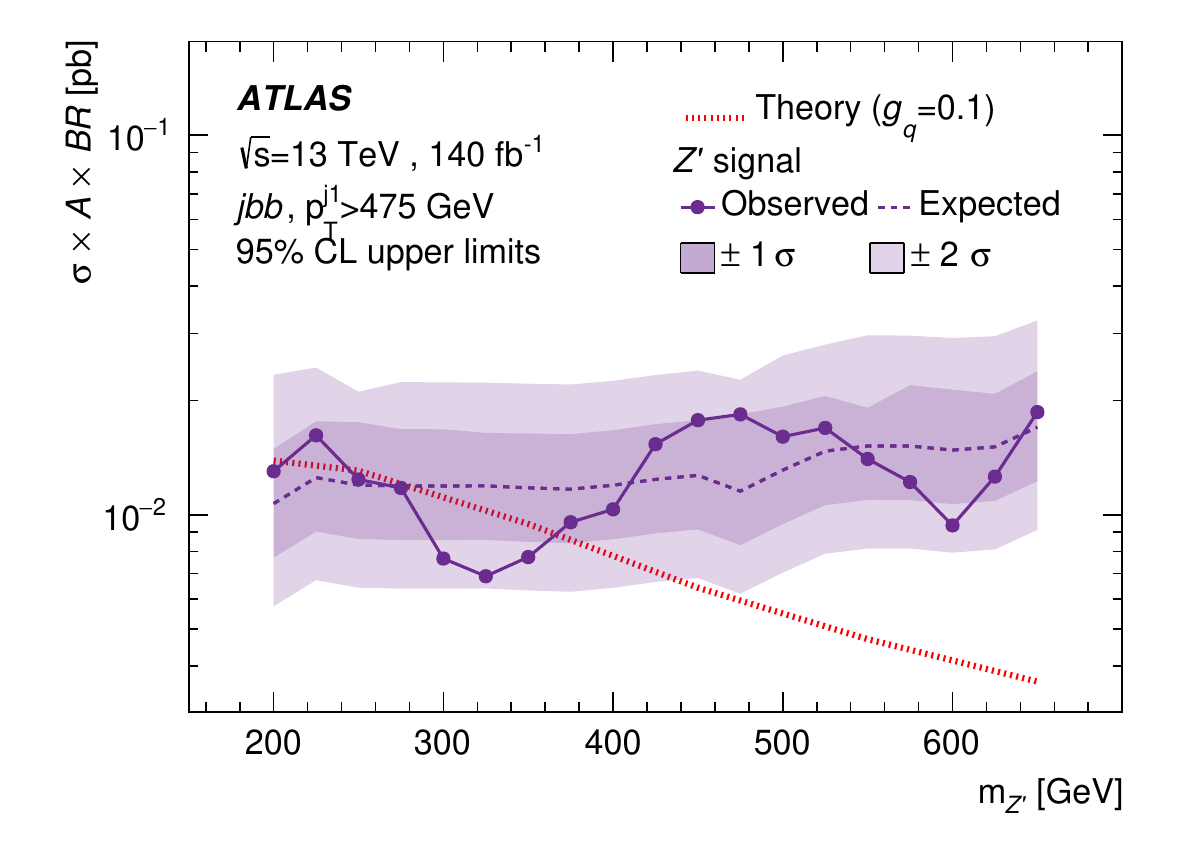}}
\caption{Expected and observed 95\% CL upper limits on the product of the cross-section, acceptance and branching ratio ($\sigma \times A \times \mathrm{BR} $) as a function of the $\Zp$ mass for the (a) \ajj, (b) \abb, (c) \jjj, and (d) \jbb~channels with the background and signal contributions modeled by a 5-parameter fit function and templates from the signal samples, respectively. Observed limits are indicated with markers and a solid line, and expected limits are indicated with a dashed line.
The shaded bands around the expected limit indicates the (darker band) $1\sigma$ and (lighter band) $2\sigma$ uncertainty range. The dotted line represents the expected $\sigma \times A \times \mathrm{BR}$ assuming $g_q=0.1$, as obtained from the \MADGRAPH signal samples.}
\label{fig:limits}
\end{figure}

\begin{figure}[tbp]
\centering
\subfigure[]{\includegraphics[width=0.48\textwidth]{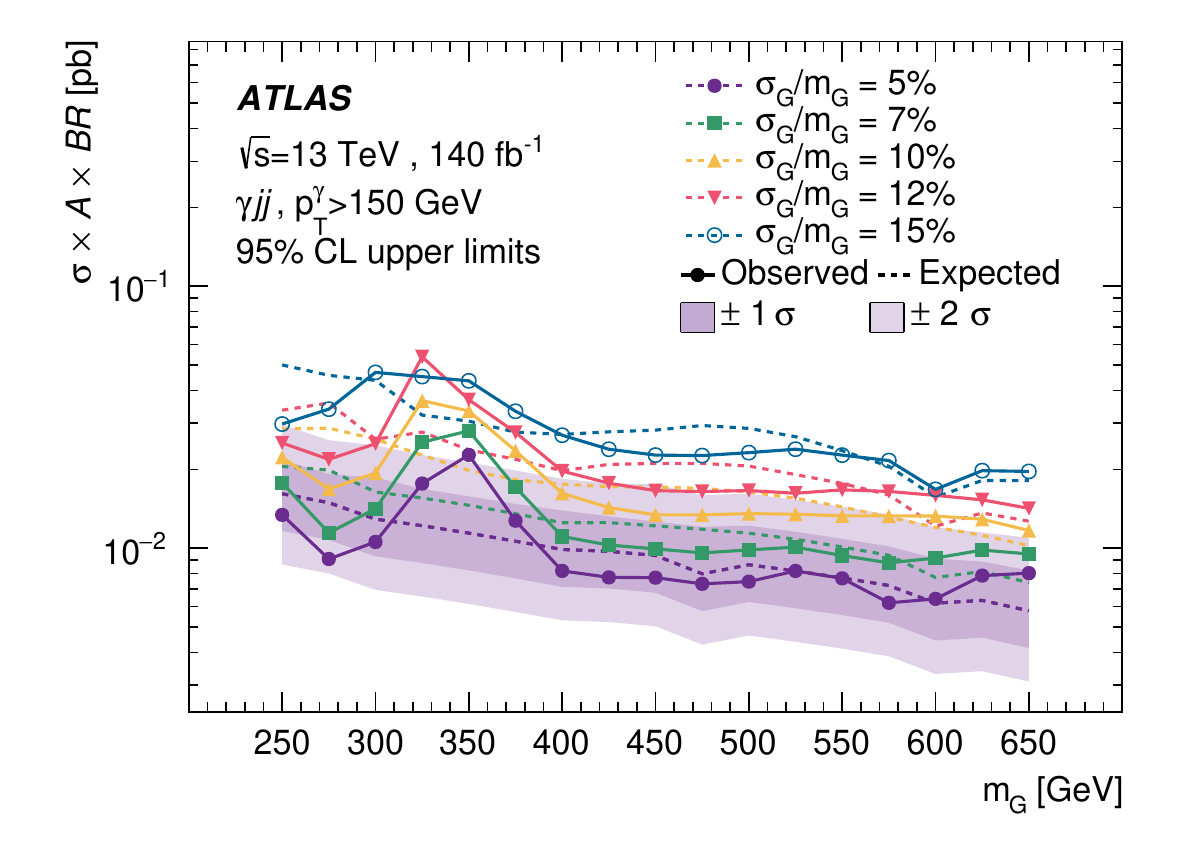}}
\subfigure[]{\includegraphics[width=0.48\textwidth]{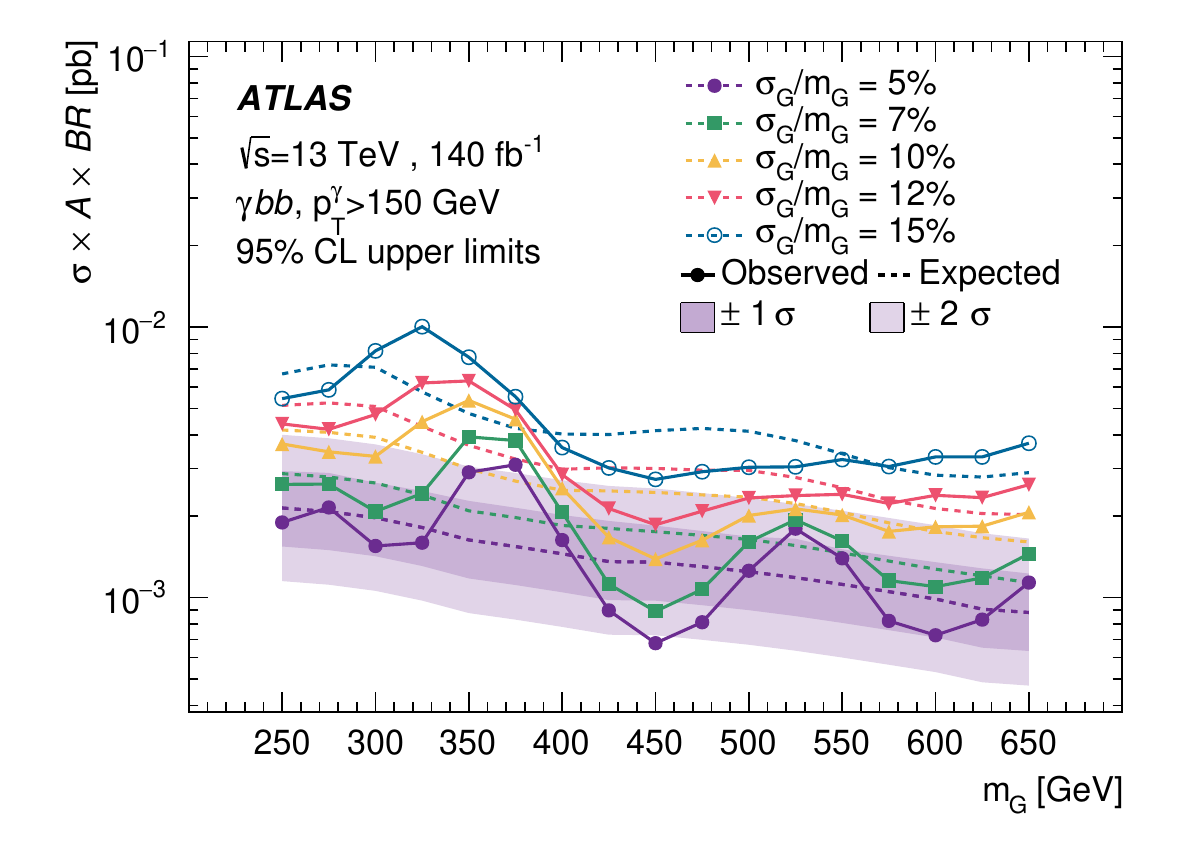}} \\
\subfigure[]{\includegraphics[width=0.48\textwidth]{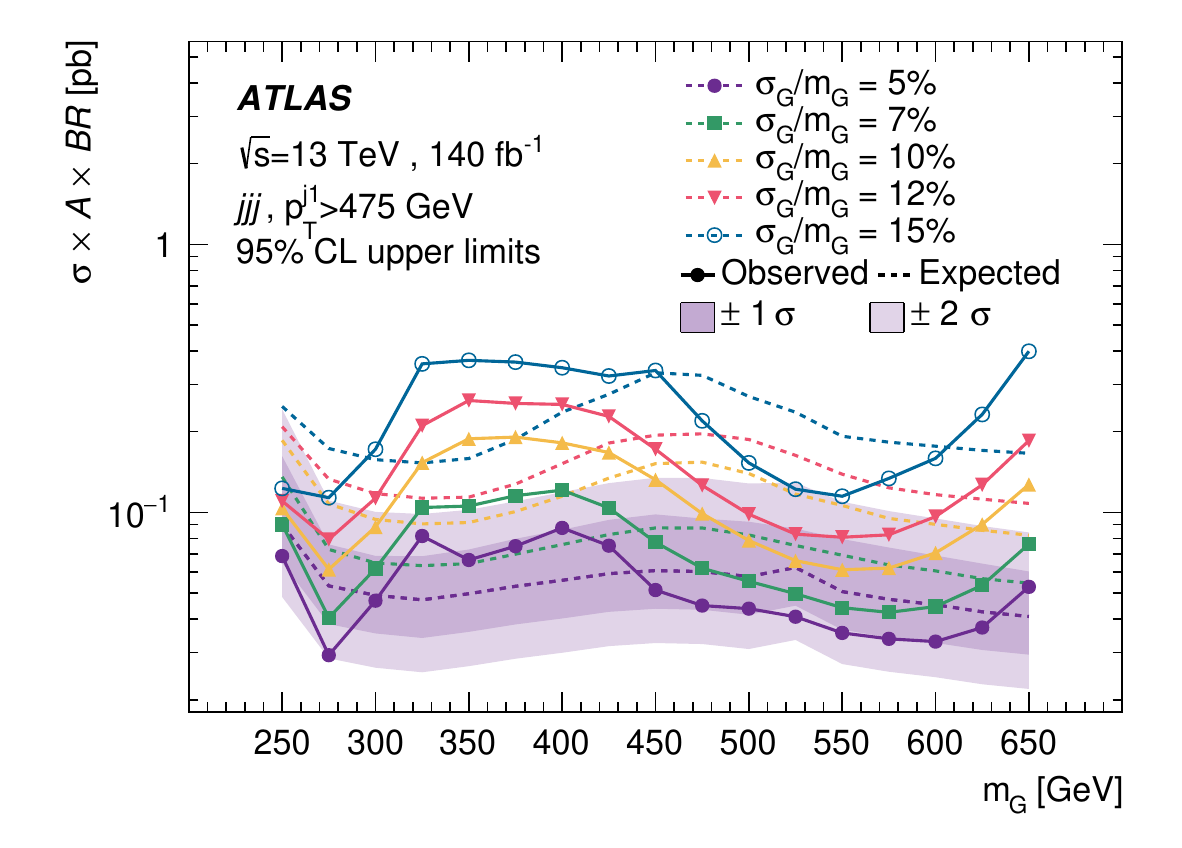}}
\subfigure[]{\includegraphics[width=0.48\textwidth]{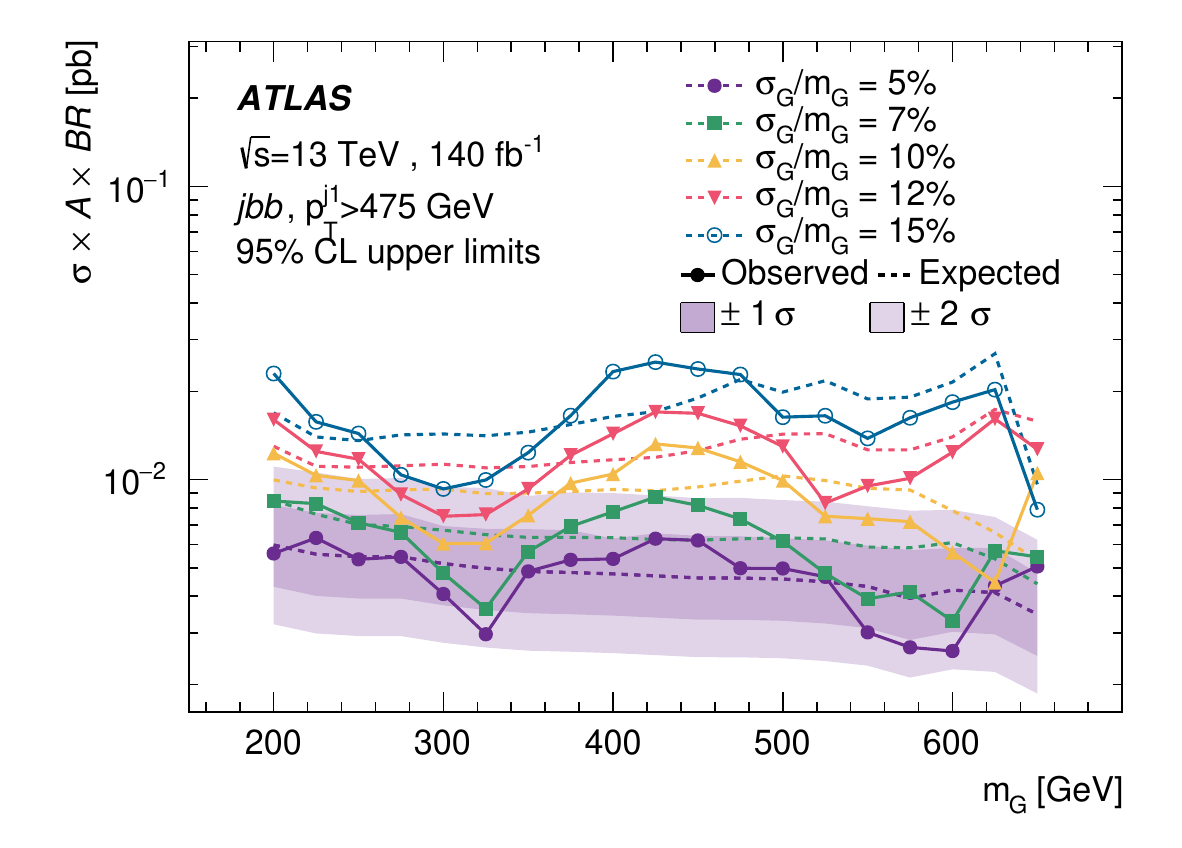}} \\
\caption{Expected and observed 95\% CL upper limits on the product of the cross-section, acceptance and branching ratio ($\sigma \times A \times \mathrm{BR} $) for the (a) \ajj, (b) \abb, (c) \jjj, and (d) \jbb~channels with the background and signal contributions modeled by a 5-parameter fit function and Gaussian distributed signal templates with a mass $m_\text{G}$ and width $\sigma_\text{G}$, respectively.
Observed limits are indicated with markers and a solid line, and expected limits are indicated with a dashed line.
Observed and expected limits corresponding to different choices of template widths (5\%, 7\%, 10\%, 12\% and 15\%) are indicated as different sets of markers or line styles, respectively.
The shaded bands around the expected limit for templates with 5\% width indicate the (darker band) $1\sigma$ and (lighter band) $2\sigma$ uncertainty range.}
\label{fig:limits:gaus}
\end{figure}

Given the relation between couplings and cross-section, the limits on the \Zp~cross-sections can be translated to contraints on the $g_q$ coupling,
by taking into account the values of the simulated sample cross-sections and the analysis acceptance.
A comparison of the limits on the $g_q$ coupling as a function of the \Zp~mass for all channels is shown in Figure~\ref{fig:limitsComparison}.
For the flavor-inclusive channels, the cross-section combines the contributions from the signal samples where the \Zp~decays are restricted to light flavors with the samples where the \Zp~decay products decay into $b$-quarks.
While the limits on $\sigma \times A \times \mathrm{BR}$ are much stronger for the \abb~and \jbb~channels than for the \ajj~and \jjj~channels,
the cross-sections and branching ratios are much lower, meaning that the limits on $g_q$ are more similar across channels.
Overall, the \jjj~channel sets the strongest limits on the $g_q$ coupling.

\begin{figure}[tbp]
\centering
\includegraphics[width=0.48\textwidth]{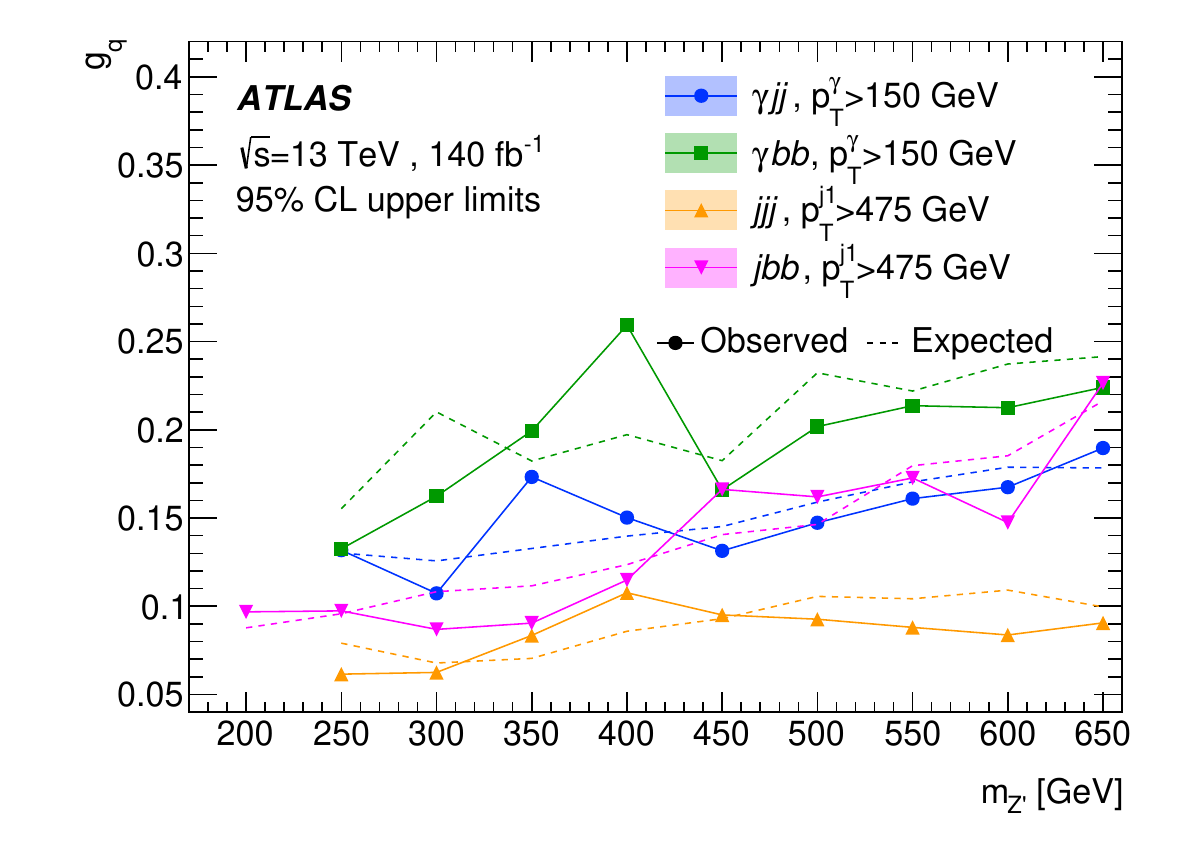}
\caption{Expected and observed 95\% CL upper limits on the $g_q$ coupling as a function of the \Zp~mass on the \ajj, \abb, \jjj, and \jbb~channels with the background and signal contributions modeled by a 5-parameter fit function and templates from the signal samples, respectively.
Observed limits are indicated with markers and a solid line, and expected limits are indicated with a dashed line.
}
\label{fig:limitsComparison}
\end{figure}

Given that there is only a small overlap in the selected events across the \ajj\ and \jjj\ categories, a combination of the two channels can further strengthen the constraints on $g_q$.
No combination of the $b$-tagged channel is considered, as there the overlap in selected events is not negligible.

Events passing both the \ajj~ and \jjj~selections are removed from the \jjj~channel,  to have two statistically independent measurements. Such double-counted events amount to $0.2\%$ of all the \jjj-selected events and their impact on the \jjj~channel fits were found to be negligible.
A combined likelihood function, which is the product of the \ajj~and \jjj~likelihood functions, is used to obtain 95\% CL upper limits on $g_q$.
In the combined likelihood, the jet and luminosity systematic uncertainties are taken as correlated among the channels. Spurious signal and PDF systematic uncertainties are instead taken as uncorrelated, as the background fit functions and production mechanisms of the two processes are independent across channels.
Photon systematic uncertainties only affect the \ajj~part of the combined likelihood.
The combination does not significantly improve the limits on $g_q$ relative to the \jjj\ channel, as shown in Figure~\ref{fig:limitsComb}. This is expected, given that the \ajj\ channel has a reduced sensitivity to the presence of signal. Both the expected and observed limits exclude coupling values down to 0.06--0.1.

\begin{figure}[tbp]
\centering
\includegraphics[width=0.48\textwidth]{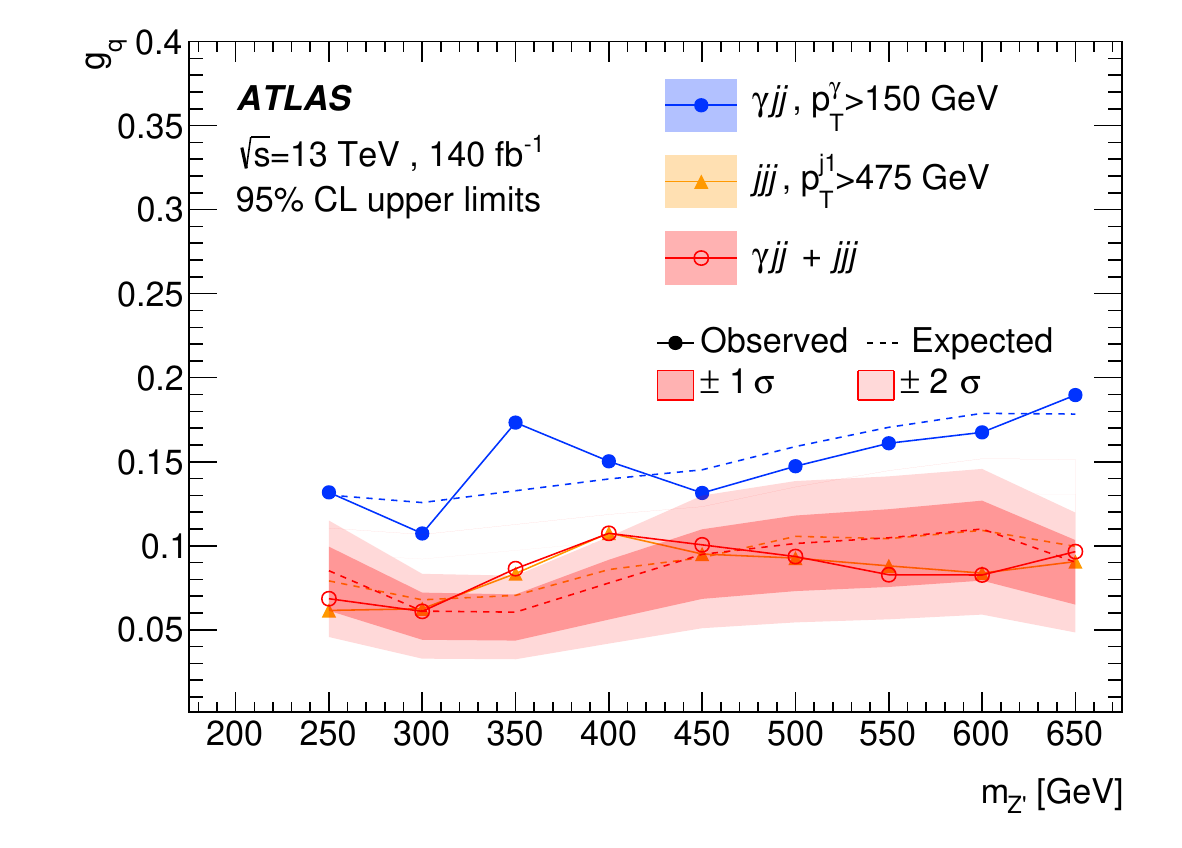}
\caption{Expected and observed 95\% CL upper limits on the $g_q$ coupling as a function of the \Zp~mass on the \ajj~channel, the \jjj~channel and the combination of the two. The background and signal contributions are modeled by a 5-parameter fit function and templates from the signal samples, respectively.
Observed limits are indicated with markers and a solid line, and expected limits are indicated with a dashed line.
The shaded bands indicate the (darker band) 1$\sigma$ and (lighter band) 2$\sigma$ uncertainty range on the combined result.
}
\label{fig:limitsComb}
\end{figure}

% End of text imported from the .//sections/results.tex input file

%
%
\FloatBarrier

\section{Conclusion}
\label{sec:conclusion}

% The next lines are included from the .//sections/conclusions.tex input file
Dijet resonances with a width up to 15\% of the mass, produced in association with a photon or jet, were searched for in 140 \ifb of LHC $pp$ collisions recorded by the ATLAS experiment at $\sqrt{s} = 13$~\TeV.
This search expands on previous similar searches by using the full Run~2 data samples, and by including the case where the initial-state radiation is a jet.
It considers the cases where no flavor requirements are placed on the resonance decay products, and the case where both of the decay products are required to be $b$-tagged,
resulting in four channels, depending on the type of initial-state radiation and the flavor requirements.
In all four channels, the observed \mjj~distribution is well-described by a smooth functional fit without contributions from such resonances.
No significant excess of events beyond the Standard Model expectation is observed, and so upper limits are set on two models: \Zp~axial-vector dark-matter mediators and Gaussian-shape signal contributions
for resonant masses between 200~\GeV~and 650~\GeV.
Relative to already published results using a similar analysis technique, this search improves the limits on the $\Zp\text{-}q\text{-}q$ coupling $g_q$ by up to 50\%,
with the most stringent limits on $g_q$ set by the \jjj~channel.
A further combination of the \jjj~and \ajj~channels does not bring a significant improvement with respect to the \jjj\ channel alone, and the limits on $g_q$ are set down to 0.06--0.1.

% End of text imported from the .//sections/conclusions.tex input file

%

%
\section*{Acknowledgements}
%

% The next lines are included from the .//acknowledgements/Acknowledgements.tex input file
%
%

%
%

We thank CERN for the very successful operation of the LHC and its injectors, as well as the support staff at
CERN and at our institutions worldwide without whom ATLAS could not be operated efficiently.

The crucial computing support from all WLCG partners is acknowledged gratefully, in particular from CERN, the ATLAS Tier-1 facilities at TRIUMF/SFU (Canada), NDGF (Denmark, Norway, Sweden), CC-IN2P3 (France), KIT/GridKA (Germany), INFN-CNAF (Italy), NL-T1 (Netherlands), PIC (Spain), RAL (UK) and BNL (USA), the Tier-2 facilities worldwide and large non-WLCG resource providers. Major contributors of computing resources are listed in Ref.~\cite{ATL-SOFT-PUB-2023-001}.

We gratefully acknowledge the support of ANPCyT, Argentina; YerPhI, Armenia; ARC, Australia; BMWFW and FWF, Austria; ANAS, Azerbaijan; CNPq and FAPESP, Brazil; NSERC, NRC and CFI, Canada; CERN; ANID, Chile; CAS, MOST and NSFC, China; Minciencias, Colombia; MEYS CR, Czech Republic; DNRF and DNSRC, Denmark; IN2P3-CNRS and CEA-DRF/IRFU, France; SRNSFG, Georgia; BMBF, HGF and MPG, Germany; GSRI, Greece; RGC and Hong Kong SAR, China; ISF and Benoziyo Center, Israel; INFN, Italy; MEXT and JSPS, Japan; CNRST, Morocco; NWO, Netherlands; RCN, Norway; MNiSW, Poland; FCT, Portugal; MNE/IFA, Romania; MESTD, Serbia; MSSR, Slovakia; ARRS and MIZ\v{S}, Slovenia; DSI/NRF, South Africa; MICINN, Spain; SRC and Wallenberg Foundation, Sweden; SERI, SNSF and Cantons of Bern and Geneva, Switzerland; MOST, Taipei; TENMAK, T\"urkiye; STFC, United Kingdom; DOE and NSF, United States of America.

Individual groups and members have received support from BCKDF, CANARIE, CRC and DRAC, Canada; CERN-CZ, PRIMUS 21/SCI/017 and UNCE SCI/013, Czech Republic; COST, ERC, ERDF, Horizon 2020, ICSC-NextGenerationEU and Marie Sk{\l}odowska-Curie Actions, European Union; Investissements d'Avenir Labex, Investissements d'Avenir Idex and ANR, France; DFG and AvH Foundation, Germany; Herakleitos, Thales and Aristeia programmes co-financed by EU-ESF and the Greek NSRF, Greece; BSF-NSF and MINERVA, Israel; Norwegian Financial Mechanism 2014-2021, Norway; NCN and NAWA, Poland; La Caixa Banking Foundation, CERCA Programme Generalitat de Catalunya and PROMETEO and GenT Programmes Generalitat Valenciana, Spain; G\"{o}ran Gustafssons Stiftelse, Sweden; The Royal Society and Leverhulme Trust, United Kingdom.

In addition, individual members wish to acknowledge support from CERN: European Organization for Nuclear Research (CERN PJAS); Chile: Agencia Nacional de Investigaci\'on y Desarrollo (FONDECYT 1190886, FONDECYT 1210400, FONDECYT 1230812, FONDECYT 1230987); China: Chinese Ministry of Science and Technology (MOST-2023YFA1605700), National Natural Science Foundation of China (NSFC - 12175119, NSFC 12275265, NSFC-12075060); Czech Republic: PRIMUS Research Programme (PRIMUS/21/SCI/017); EU: H2020 European Research Council (ERC - 101002463); European Union: European Research Council (ERC - 948254, ERC 101089007), Horizon 2020 Framework Programme (MUCCA - CHIST-ERA-19-XAI-00), European Union, Future Artificial Intelligence Research (FAIR-NextGenerationEU PE00000013), Italian Center for High Performance Computing, Big Data and Quantum Computing (ICSC, NextGenerationEU); France: Agence Nationale de la Recherche (ANR-20-CE31-0013, ANR-21-CE31-0013, ANR-21-CE31-0022), Investissements d'Avenir Labex (ANR-11-LABX-0012); Germany: Baden-Württemberg Stiftung (BW Stiftung-Postdoc Eliteprogramme), Deutsche Forschungsgemeinschaft (DFG - 469666862, DFG - CR 312/5-2); Italy: Istituto Nazionale di Fisica Nucleare (ICSC, NextGenerationEU); Japan: Japan Society for the Promotion of Science (JSPS KAKENHI JP21H05085, JSPS KAKENHI JP22H01227, JSPS KAKENHI JP22H04944, JSPS KAKENHI JP22KK0227); Netherlands: Netherlands Organisation for Scientific Research (NWO Veni 2020 - VI.Veni.202.179); Norway: Research Council of Norway (RCN-314472); Poland: Polish National Agency for Academic Exchange (PPN/PPO/2020/1/00002/U/00001), Polish National Science Centre (NCN 2021/42/E/ST2/00350, NCN OPUS nr 2022/47/B/ST2/03059, NCN UMO-2019/34/E/ST2/00393, UMO-2020/37/B/ST2/01043, UMO-2021/40/C/ST2/00187, UMO-2022/47/O/ST2/00148); Slovenia: Slovenian Research Agency (ARIS grant J1-3010); Spain: BBVA Foundation (LEO22-1-603), Generalitat Valenciana (Artemisa, FEDER, IDIFEDER/2018/048), Ministry of Science and Innovation (MCIN \& NextGenEU PCI2022-135018-2, MICIN \& FEDER PID2021-125273NB, RYC2019-028510-I, RYC2020-030254-I, RYC2021-031273-I, RYC2022-038164-I), PROMETEO and GenT Programmes Generalitat Valenciana (CIDEGENT/2019/023, CIDEGENT/2019/027); Sweden: Swedish Research Council (VR 2018-00482, VR 2022-03845, VR 2022-04683, VR grant 2021-03651), Knut and Alice Wallenberg Foundation (KAW 2017.0100, KAW 2018.0157, KAW 2018.0458, KAW 2019.0447, KAW 2022.0358); Switzerland: Swiss National Science Foundation (SNSF - PCEFP2\_194658); United Kingdom: Leverhulme Trust (Leverhulme Trust RPG-2020-004), Royal Society (NIF-R1-231091); United States of America: U.S. Department of Energy (ECA DE-AC02-76SF00515), Neubauer Family Foundation.

%
%

% End of text imported from the .//acknowledgements/Acknowledgements.tex input file

%
%
%
%
%
%

%
%

%
%
%
\printbibliography

\clearpage
% ATLAS Collaboration author list
% Reference date of EXOT-2019-01 is 2023-12-15
% Author list last updated on date 28-JUN-24
% Data extracted on 28-Jun-2024 for paper reference EXOT-2019-01
% at 2:00am
 
\begin{flushleft}
\hypersetup{urlcolor=black}
{\Large The ATLAS Collaboration}

\bigskip

\AtlasOrcid[0000-0002-6665-4934]{G.~Aad}$^\textrm{\scriptsize 103}$,
\AtlasOrcid[0000-0001-7616-1554]{E.~Aakvaag}$^\textrm{\scriptsize 16}$,
\AtlasOrcid[0000-0002-5888-2734]{B.~Abbott}$^\textrm{\scriptsize 121}$,
\AtlasOrcid[0000-0002-1002-1652]{K.~Abeling}$^\textrm{\scriptsize 55}$,
\AtlasOrcid[0000-0001-5763-2760]{N.J.~Abicht}$^\textrm{\scriptsize 49}$,
\AtlasOrcid[0000-0002-8496-9294]{S.H.~Abidi}$^\textrm{\scriptsize 29}$,
\AtlasOrcid[0009-0003-6578-220X]{M.~Aboelela}$^\textrm{\scriptsize 44}$,
\AtlasOrcid[0000-0002-9987-2292]{A.~Aboulhorma}$^\textrm{\scriptsize 35e}$,
\AtlasOrcid[0000-0001-5329-6640]{H.~Abramowicz}$^\textrm{\scriptsize 152}$,
\AtlasOrcid[0000-0002-1599-2896]{H.~Abreu}$^\textrm{\scriptsize 151}$
% ...
\bigskip
\\

% ...
$^{16}$Department for Physics and Technology, University of Bergen, Bergen; Norway.\\
% ...
$^{29}$Physics Department, Brookhaven National Laboratory, Upton NY; United States of America.\\
% ...
$^{35}$$^{(a)}$Facult\'e des Sciences Ain Chock, R\'eseau Universitaire de Physique des Hautes Energies - Universit\'e Hassan II, Casablanca;$^{(b)}$Facult\'{e} des Sciences, Universit\'{e} Ibn-Tofail, K\'{e}nitra;$^{(c)}$Facult\'e des Sciences Semlalia, Universit\'e Cadi Ayyad, LPHEA-Marrakech;$^{(d)}$LPMR, Facult\'e des Sciences, Universit\'e Mohamed Premier, Oujda;$^{(e)}$Facult\'e des sciences, Universit\'e Mohammed V, Rabat;$^{(f)}$Institute of Applied Physics, Mohammed VI Polytechnic University, Ben Guerir; Morocco.\\
% ...
$^{44}$Physics Department, Southern Methodist University, Dallas TX; United States of America.\\
% ...
$^{49}$Fakult\"{a}t Physik , Technische Universit{\"a}t Dortmund, Dortmund; Germany.\\
% ...
$^{55}$II. Physikalisches Institut, Georg-August-Universit\"{a}t G\"ottingen, G\"ottingen; Germany.\\
% ...
$^{103}$CPPM, Aix-Marseille Universit\'e, CNRS/IN2P3, Marseille; France.\\
% ...
$^{121}$Homer L. Dodge Department of Physics and Astronomy, University of Oklahoma, Norman OK; United States of America.\\
% ...
$^{151}$Department of Physics, Technion, Israel Institute of Technology, Haifa; Israel.\\
$^{152}$Raymond and Beverly Sackler School of Physics and Astronomy, Tel Aviv University, Tel Aviv; Israel.\\
% ...
$^{a}$ Also Affiliated with an institute covered by a cooperation agreement with CERN.\\
$^{b}$ Also at An-Najah National University, Nablus; Palestine.\\
$^{c}$ Also at Borough of Manhattan Community College, City University of New York, New York NY; United States of America.\\
$^{d}$ Also at Center for High Energy Physics, Peking University; China.\\
$^{e}$ Also at Center for Interdisciplinary Research and Innovation (CIRI-AUTH), Thessaloniki; Greece.\\
$^{f}$ Also at Centro Studi e Ricerche Enrico Fermi; Italy.\\
% ...

\end{flushleft}

% Created with Glance <Atlas.Glance@cern.ch>

\end{document}